\documentclass{WileyMSP-template}

\usepackage{ragged2e}
\AtBeginDocument{
  \justifying
  \setlength{\parindent}{0pt}
}

\begin{document}

\pagestyle{fancy}
\rhead{\includegraphics[width=2.5cm]{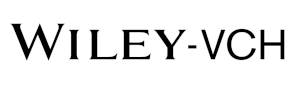}}

\title{Cryogenic Voltage Control of Magnetism in Silicon-Integrated \newline SrTiO$_3$/Fe Heterostructures}

\maketitle

\author{Stijn Reniers$^{ 1}$*}
\author{Emile Fourneau$^{ 2}$}
\author{Andries Boelen$^{ 3,4}$}
\author{Xing-Jian Liu$^{ 5}$}
\author{Ekaterina Gorokh$^{ 1}$}
\author{Lukas Nulens$^{ 1}$}
\author{Vivek Kumar$^{ 1}$}
\author{Luca Ceccon$^{ 3,4}$}
\author{Christian Haffner$^{ 4}$}
\author{Clement Merckling$^{ 3,4}$}
\author{Jun-Yi Ge$^{ 5}$*}
\author{Bertrand Dup\'e$^{ 6}$}
\author{Alejandro V. Silhanek$^{ 2}$}
\author{Kristiaan Temst$^{ 1,4}$}
\author{Joris Van de Vondel$^{ 1}$*}


\dedication{}
\begin{affiliations}

$^{1}$ Quantum Solid-State Physics, Department of Physics and Astronomy, KU Leuven, Celestijnenlaan 200D, Leuven B-3001, Belgium

$^{2}$Experimental Physics of Nanostructured Materials, Q-MAT research unit, Department of Physics, Université de Liège, Liège B-4000, Belgium

$^{3}$Department of Materials Engineering (MTM), KU Leuven, B-3001 Leuven, Belgium

$^{4}$Imec, Kapeldreef 75, Leuven, Belgium

$^{5}$Materials Genome Institute, Shanghai University, Shanghai 200444, China

$^{6}$TOM research group, Q-MAT research unit, Université de Liège, Liège B-4000, Belgium

*stijn.reniers@kuleuven.be
*joris.vandevondel@kuleuven.be
*junyi\_ge@t.shu.edu.cn
\end{affiliations}



\begin{abstract}
Cryogenic electronics forms a rapidly emerging research domain for high-performance and power-efficient computing applications. Incorporating nanomagnetic components in cryogenic circuitry adds highly valuable functionality, facilitating downscaling, reducing energy consumption and introducing time-reversal symmetry breaking. Furthermore, low-temperature environments enhance magnetic stability and switching efficiency at nanoscale dimensions, reinforcing the potential of cryogenic nanomagnets. To fully leverage these opportunities, magnetic control schemes require alternative options to current-based writing, which is the main bottleneck regarding power consumption and downscaling. In this regard, voltage-based gating of the magnetic state could drastically enhance operational efficiency and integration density. In this work, we investigate cryogenic Voltage Control of Magnetism (VCM) in epitaxial SrTiO$_3$/Fe thin film heterostructures on a CMOS compatible Si substrate. We demonstrate and quantify voltage-controlled modifications of the magnetic domain structure, consistent with electric field-controlled magnetic anisotropy at the Fe/SrTiO$_3$ interface. These findings provide a viable material system for the development of next-generation magnetic domain-based devices for classical and quantum computing.

\end{abstract}


\section{Introduction}\label{Intro}

Modern computing technology faces a growing demand for high-performance and low-power operation, while maintaining scalable integration \cite{MotivationComputingEff1, WangExponentialScalingBeyondCMOS17, FrankMPEfficiency25}. Cryogenic electronics has emerged as a powerful approach, harnessing the unique low-temperature behavior of semiconducting, magnetic and superconducting materials to introduce new device concepts for classical and quantum computing. While enabling enhanced switching speeds and computational power, cryogenic electronics can maintain low power dissipation and high circuit reliability \cite{CryogenicLowPowerElectronics, CryogenicMemory, BashirCryogenic25}. In view of these advancements, nanomagnetic devices operating at low temperatures could offer unique additional functionality.  When combined with superconducting circuitry, spin-polarized, dissipationless supercurrents can be established by controlling the magnetic configuration in spin-valve structures or magnetic Josephson junctions \cite{RyazanovReview, BirgeMagneticJunctions, RobinsonJunctionPi}. Such components are extremely useful to implement ultralow power superconducting spintronics \cite{Superspintron, YangSuperspintronics21, CaiSuperspintronics23}, Rapid Single Flux Quantum (RSFQ) logic \cite{RyazanovReview, MagneticJJforRSFQ_Vernik, SFMemorycell_RSFQ_Nevirkovets23} and superconducting qubits \cite{YamashitaFerrotransmon, Kim2024} with reduced device footprint and nonvolatile operation. In addition, cryogenic environments, and the consequent reduction of thermal fluctuations, facilitate scalability \cite{Cowburnsizelimit, WellerSizelimit, ThompsonSizelimit, SkumryevSizelimit2003}, domain wall stability and power efficiency of the magnetic switching process \cite{CryomagneticAnisotropy25, ThermalSTTCryogenic, Veiga23_CryoSTT, pMTJ_Cryo_Lang_APL_20, Rowlands2019, LTMagneticMemoryLang2020, CrySpinHallMemory}, further strengthening the impact of nanomagnetic integration at low temperatures.

\vspace{10 pt}

To be viable for high-end application, the combination of high computational performance and density with limited intrinsic dissipation has to exceed the cost of cryogenic cooling and the power consumption of external cabling outside the cryogenic environment. Additionally, the circuit scalability aspect remains quite poorly integrated as long as current bias lines, with large inherent footprint, are required to control the magnetic state. Significant progress in the aforementioned aspects could be obtained by reconsideration of the magnetic state control scheme from resistive switching towards capacitive gating, similar to the CMOS device concept.  \textit{Voltage Control of Magnetism (VCM)} refers to the field of research in which the magnetic configuration of thin films can be controlled by electric fields, applied across a dielectric gate-oxide \cite{SongReview, VCMAreviewNozakiMiwa, ECMAReviewFert, VCMAReviewMishra, VcontrolledMagnonics}. VCM is considerably less demanding in terms of power consumption and circuit footprint than conventional current-based control methods. Physical manifestations of VCM can occur by different \textit{magneto-electric} mechanisms that establish a coupling between the polarization of the dielectric and the magnetic order in the ferromagnetic layer, i.e., strain transfer (\textit{mechanical}), ionic transport (\textit{chemical}) or interfacial charge accumulation and orbital hybridization (\textit{electronic}). 

\vspace{10 pt}

With MRAM technology as the most direct application perspective, the majority of spintronics and VCM research is directed towards room-temperature implementations. However, given the unique potential of gate-controlled nanomagnetic components in low temperature electronics and quantum technology, it is of significant strategic interest to investigate VCM specifically in the cryogenic regime. The successful demonstration of a CMOS-friendly magnetic material system, that combines nanoscale dimensionality with voltage control of the magnetic configuration at cryogenic temperatures, could strongly accelerate device implementations that fully exploit the potential of high-performance cryogenic electronics and spintronics. 

\vspace{10 pt}

Substantial VCM-effects have been reported earlier on bulk ferroelectric \cite{VCMACoSTO, CoPtPMNPTSRT,CoPtPMNPTDeterministic, Bauer2015, Lahtinen2012, Li2017, QinSpinWavegating}, electrolytic \cite{Quintana_Sort, ZhaoIonicLiquid} and thin film dielectric \cite{Shukla2018_Fe_Pd_MgO, NakamuraVCM_FeMgO, NozakiVCMMgO, VCMASTOVermeulen} gate oxides. The coupling mechanisms in bulk systems are almost exclusively strain-mediated or ionic in nature, making them hardly scalable, difficult to integrate in CMOS technology and usually incompatible with cryogenic environments. In the context of high-density, cryogenic operation, a solid-state and purely \textit{electronic} VCM effect is required, i.e., mediated solely by charge accumulation and depletion at the magnet-dielectric interface upon polarization of the dielectric. This is also known as \textit{Voltage Control of Magnetic Anisotropy (VCMA)}, since it relies on modification of the magnetic anisotropy component at the ferromagnet/dielectric interface. High-permittivity, thin film gate dielectrics, combined with an ultrathin ferromagnetic metal in a crystalline heterostructure, are ideal for this purpose. In this regard, the SrTiO$_3$/Fe heterostructure presents a promising alternative to state-of-the art implementations, which focus mainly on MgO as the gate oxide. High-quality SrTiO$_3$ thin films provide an excellent template for the growth of epitaxial ferromagnetic films. Moreover, SrTiO$_3$ exhibits a large tunability window in dielectric and even ferroelectric properties by varying the temperature, stoichiometry, film thickness or strain state. Previous experimental work on the SrTiO$_3$/Fe system has focused mainly on structural parameters \cite{Cho_2004, Kamaratos_2008, Chien12}, interfacial properties \cite{Catrou18} and the Fe magnetic behavior \cite{Cho_2003} on bulk substrates, demonstrating good crystalline quality and robust ferromagnetism of the Fe film. A CMOS-friendly, thin film implementation of the epitaxial SrTiO$_3$/Fe system would provide a significant reference material for cryogenic VCM, with several unique advantages. Firstly, thermally-assisted leakage currents through the SrTiO$_3$ gate oxide are substantially reduced at low temperature, enhancing power-efficiency and preserving the capacitive nature of the gate effect. Secondly, the dielectric permittivity $\epsilon_r$ of SrTiO$_3$ is expected to be enhanced at low temperatures \cite{Muller79, STO_permittivity2, STO_permittivity_3}, where it approaches a ferroelectric instability that depends on composition, electric field and strain state \cite{ulrich2025engineering,Kamal_STO_permittivity}. Since for an interfacial, charge-mediated effect, the impact of voltage gating scales with the interfacial polarization charge $P = \epsilon_r \epsilon_0 E$ \cite{VCMAreviewNozakiMiwa, VCMASTOVermeulen}, the low-temperature permittivity enhancement of SrTiO$_3$ provides a advantage in operation. Furthermore, any ionic diffusion mechanism can be excluded, as such thermally-activated processes drastically slow down in a cryogenic environment. This allows to assume a purely electronic effect. Recent theoretical works on the interfacial magnetic anisotropy \cite{STOFeMCADFT} and its expected voltage control \cite{STOFeVCMADFT} predict a strong perpendicular magnetic anisotropy at the SrTiO$_3$/Fe interface and a high voltage-tunability, quantified by the VCMA coupling coefficient $\xi$. Successful experimental realization of voltage-controlled magnetism in the cryogenic regime, consistent with an established theoretical framework, would put the SrTiO$_3$/Fe system forward as a prominent gate-controlled nanomagnetic material stack for integration in high-performance, low-power spintronics and scalable quantum circuitry.   

\vspace{10 pt}

In this work, we fabricate epitaxial SrTiO$_3$/Fe/Ag thin film heterostructures, deposited by Molecular Beam Epitaxy (MBE) on a conductive, highly $<$p++$>$ doped silicon substrate. The epitaxial quality of the heterostructure is expected to enhance electronic magneto-electric coupling at the SrTiO$_3$/Fe interface and allow for a better comparison with first-principles calculations performed in this work and in other theoretical works \cite{STOFeMCADFT, STOFeVCMADFT}. To maximize the susceptibility of the Fe magnetic state to interfacial voltage-control, the total effective magnetic anisotropy, determining the equilibrium spin orientation in the Fe film, has to be minimal. This requires careful balancing of the interfacial and magnetostatic volume anisotropies, which can be achieved by thickness control of the Fe film in the ultrathin regime. By experimentally establishing the thickness dependence of the effective magnetic anisotropy energy, we extrapolate the magnetic anisotropy component at the SrTiO$_3$/Fe interface and identify the thickness of the Spin Reorientation Transition (SRT) at $t_{Fe} = 0.7$ nm. At this thickness, the effective magnetic anisotropy is minimal, as the positive (perpendicular) interfacial anisotropy component compensates the negative (in-plane) magnetostatic one. Selecting the thickness $t_{Fe} = 0.8$ nm, close to the SRT, we obtain a magnetic system with reduced intrinsic stability of the equilibrium orientation. At cryogenic temperatures, the magnetic state near the SRT is stabilized against thermal fluctuations and exhibits high susceptibility to changes in the interfacial magnetic anisotropy component. We demonstrate a clear and consistent effect of back-gate voltage on the magnetotransport properties and visualize its impact on the magnetic domain structure using Magneto-Optic Kerr Effect (MOKE) measurements. We provide a quantitative estimate of the VCM effect and a microscopic understanding based on micromagnetic simulations. Our results show that the SrTiO$_3$/Fe thin film heterostructure forms an excellent candidate for low-power, scalable and CMOS friendly cryogenic devices based on voltage-tunable magnetic configurations. 

\section{Results and discussion}

\subsection{Deposition and structural characterization} \label{DepoAnd Structural}
SrTiO$_3$(100 nm)/Fe($t_{Fe}$)/Ag(4 nm) heterostructures were grown by Molecular Beam Epitaxy (MBE) on a low-resistive, (001)-oriented silicon substrate ($\rho < 0.010$ $\Omega\cdot$cm). Epitaxial thin film SrTiO$_3$ was deposited on a 200 mm wafer, under conditions optimized to achieve high crystalline quality, smooth and well‑defined surface \cite{BoelenSTO25}. Details of the deposition procedure can be found in the Experimental Methods section. Since charge‑mediated voltage control of magnetic anisotropy is known to be highly sensitive to atomic‑scale roughness, interfacial disorder, and chemical inhomogeneity, particular emphasis was placed on controlling the SrTiO$_3$ surface termination and topography prior to ferromagnetic film deposition. To this end, the wafer was cleaved in smaller samples of 2 x 2 cm each, which were chemically treated to remove excess SrO on the film surface, and subsequently annealed in oxygen to improve the film crystallinity and surface quality (Supplementary Figure S1a). The annealed samples were then loaded in the UHV chamber of a different MBE system for the metallic overlayer deposition. Fe nominal thicknesses $t_{Fe}$ = 0.7, 0.8, 1.3, 2.2 and 4 nm were grown by varying the opening time of the material crucible shutter at the same deposition rate, calibrated by quartz crystal resonator. The deposition procedure was adapted to promote the structural continuity of the Fe layer, avoiding individual island morphology \cite{Fenanocrystals} (Supplementary Figure S1b). Our growth strategy leads to epitaxial, continuous ferromagnetic Fe even in the sub-nm regime, which is non-trivial given the tendency for Volmer-Weber growth in the first monolayers \cite{Cho_2004, Chien12}. A 4 nm Ag capping layer was deposited at room temperature to promote perpendicular magnetic anisotropy (PMA) at the top Fe interface \cite{Fe_Ag_Anisotropies}. Finally, a 3 nm Au protective capping layer was deposited to prevent oxidation of the ferromagnetic Fe/Ag bilayer. For samples that required longer transport and storage times (used in the TEM/EDX, MFM and FMR measurements), and were therefore more susceptible to oxidation, a 10 nm Al capping layer was used instead of the 3 nm Au cap. 

\begin{figure*}[h!]
    \centering
    \includegraphics[width=6.5 in]{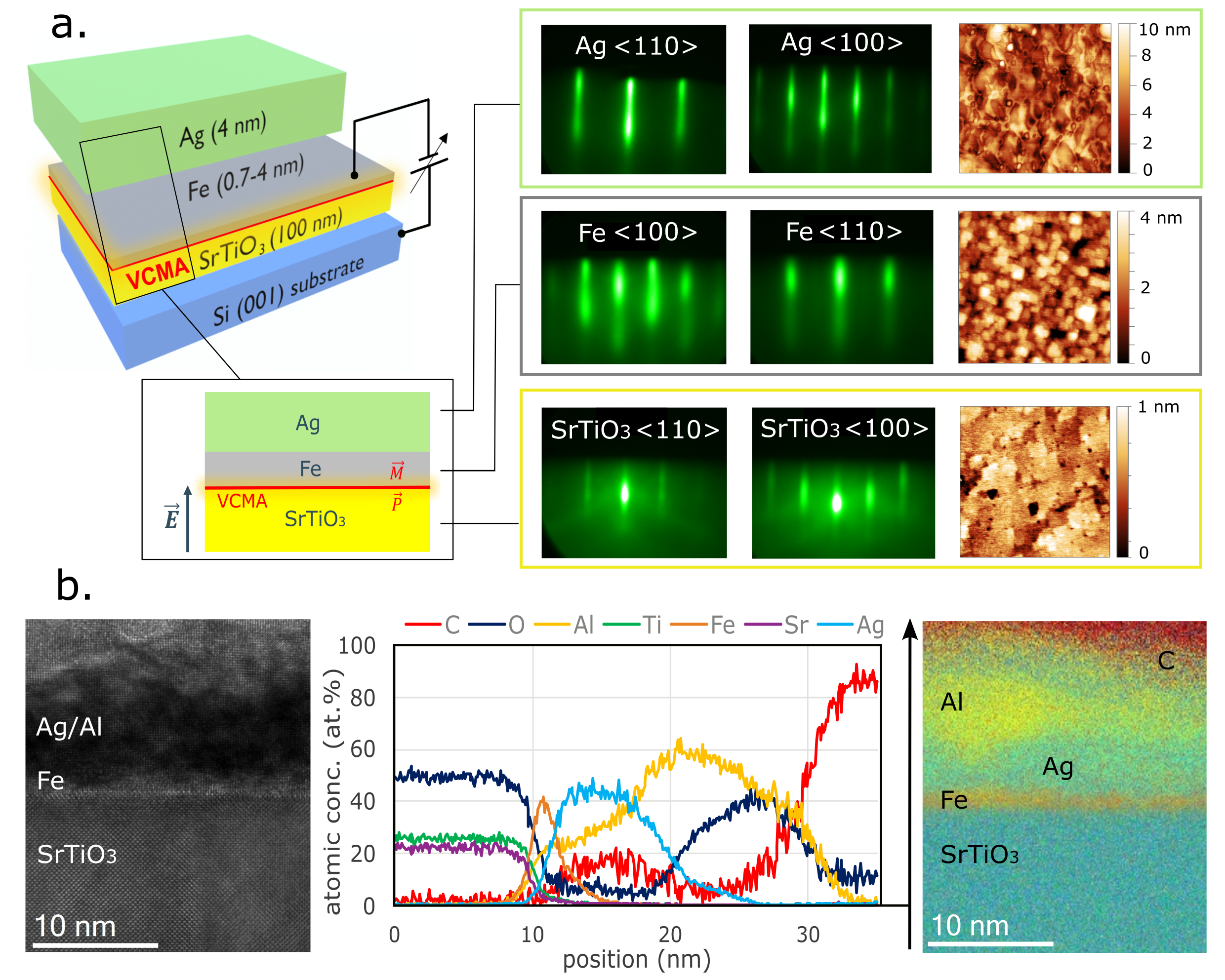}
    \caption{(a) Schematic representation of the sample stack, with the interfacial nature of electronic VCM (VCMA) indicated. For each layer surface, RHEED-recordings (during deposition) and AFM scans (after deposition, scan area 1 x 1 $\mu$m$^2$) are shown in separate insets. (b) TEM (left panel) and EDX (middle and right panel) measurements of the sample with Fe thickness 0.8 nm. The normalized integral of the EDX compositional map is plotted in the middle panel along the black arrow direction to obtain the relative elemental profile of the stack.}\label{fig:structure}
\end{figure*}
\vspace{10pt}
The system of interest is schematically shown in Figure \ref{fig:structure}a. RHEED-images of the individual film surfaces, recorded along the azimuthal directions $<$100$>$ and $<$110$>$ during the deposition, are shown in the insets of each layer, accompanied by AFM topography images that were recorded outside of the UHV environment. The nominally deposited Fe film thickness for this surface characterization was 0.8 $\pm$ 0.1 nm. Well-defined RHEED-streaks indicate the good crystalline quality of each layer surface. The intensity modulations do point out that the layers are not perfectly flat, yet the absence of spotty features implies structural continuity rather than island growth \cite{RHEED_theory}. AFM topography scans show that the SrTiO$_3$ surface prior to deposition is flat, with atomic steps of sub-nm height. To probe the Ag and Fe surface topography, two samples were deposited under identical conditions, without Al and Ag/Al capping layer, respectively. Although the ambient measurement conditions lead to oxidation and related volumetric expansion of the surface \cite{Dunn1926TheHT}, one can still discern the Fe surface structure of cubic shaped terraces, originating from the coalescence of individual islands during the Fe annealing step (Supplementary Figure S1b). As will be shown in section \ref{Magnetometry}, Fe maintains robust ferromagnetism in this sub-nm regime. The Ag capping layer is rough, and consists of partially coalesced islands, leaving a fraction of the Fe layer exposed to the protective capping. The surface topography and inhomogeneous top interface of the Fe layer have a significant impact on magnetic- and transport properties, as will be discussed in sections \ref{Magnetometry} and \ref{transport}. A cross-sectional Transmission Electron Microscopy (TEM) view of the heterostructure, shown in the left panel of Figure \ref{fig:structure}b, confirms the crystalline quality of the heterostructure and the continuity of the Fe layer. Energy-Dispersive X-ray (EDX) measurements were performed to acquire the relative elemental distribution among the individual layers. The compositional distribution near the SrTiO$_3$/Fe/Ag interface region is plotted in Figure \ref{fig:structure}b on the right hand side, with the integrated line profile in the middle section (full stack in Supplementary Figure S2a,b). The Fe nominal thickness is computed in Supplementary Information section 1 (Figure S2c) by integration of the Fe EDX profile, and found to be 1.0 nm, 25\% larger than the value of 0.8 $\pm$ 0.1 nm, calibrated by the quartz crystal. However, we still rely on the quartz calibration, since EDX can overestimate layer thickness in the sub-nm regime. Furthermore, it can be observed that there is intermixing of the Ag and Al capping layers and apparent intermixing at the Fe/Ag interface. This originates mainly from the terraced topography of both Fe and Ag, allowing different elements to occupy the vacant sites between the terraces, but interfacial alloying is not excluded. The rather inhomogenous top interface can induce significant local variations in the magnetic anisotropy, as discussed in sections \ref{Magnetometry}-\ref{spatially_res}.  However, TEM and EDX data show that the continuous Fe film maintains solid compositional integrity and that the Fe/SrTiO$_3$ interface is well-defined. The epitaxial quality of the SrTiO$_3$ thin-film and, in particular, the structural and chemical abruptness of the SrTiO$_3$/Fe interface, play a decisive role in enabling voltage control of magnetism in our material stack.

\subsection{Macroscopic magnetic characterization: Spin reorientation transition} \label{Magnetometry}

As mentioned in the introductory section, the charge-mediated VCM mechanism relies on electric field-induced modifications of the magnetic anisotropy component at the SrTiO$_3$/Fe interface. To maximize the impact of such interfacial effect on the global magnetic configuration, the total effective magnetic anisotropy energy of the Fe film should be minimal. This can be achieved by controlling the magnetic film thickness. The approach based on thickness-mediated balancing of competing volume and interface anisotropies is encompassed in the linear \textit{Gradmann} ansatz \cite{SRTAnsatzOepen}. 

\begin{equation}
    K_{eff}\cdot t_{Fe} = K_{v}\cdot t_{Fe}  + (K_{s,Fe/SrTiO_3} + K_{s,top}) \label{Gradmann} .
\end{equation}

Here, $K_{eff}$ is the effective magnetic anisotropy energy density, i.e., the volume-averaged sum of all local surface ($K_s$) and volumetric ($K_v$, magnetostatic and magnetocrystalline) anisotropy contributions in the Fe film. The optimal Fe thickness should achieve a balance of the volumetric film anisotropy (favoring in-plane magnetization) with the interface anisotropies (favoring out-of-plane magnetization).

\begin{figure*}[h]
    \centering
    \includegraphics[width=7.5 in]{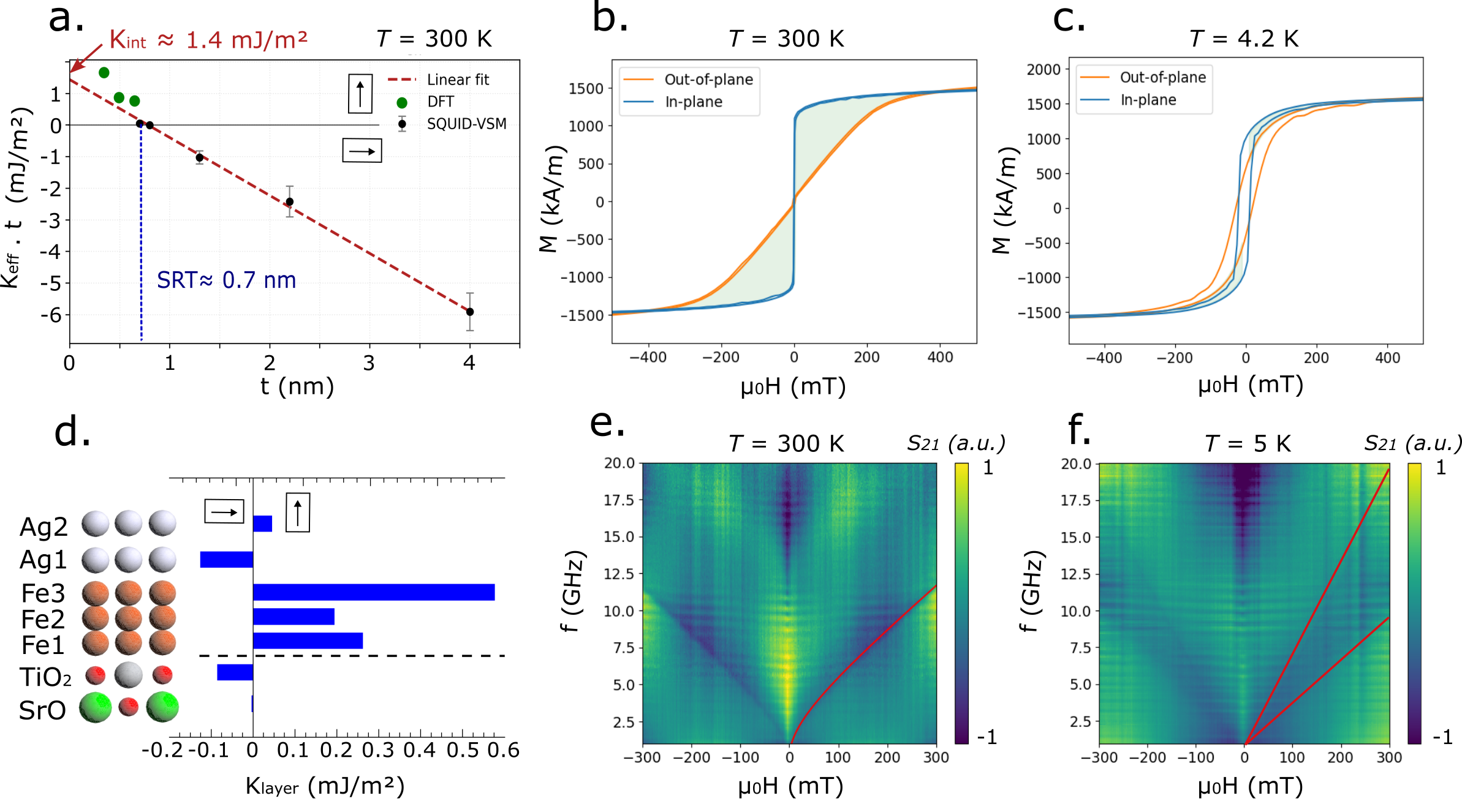}
    \caption{(a) Thickness dependence of the effective areal magnetic anisotropy energy density $K_{eff}\cdot t_{Fe}$ at $T$ = 300 K, following a linear trend consistent with equation (\ref{Gradmann}). $K_{eff}$ was obtained from magnetization hysteresis loop analysis of samples with different $t_{Fe}$, shown here for $t_{Fe}$ = 0.8 nm at (b) 300 K and (c) 4.2 K. The green points are computed from DFT calculations on continuous Fe layers of 2, 3 and 4 monolayers. (d) DFT model of a continuous 3 monolayer Fe system, with layer-resolved anisotropy contributions. FMR resonance spectra for $t_{Fe}$ = 0.8 nm were recorded at $T$ = (e) 300 K and (f) 5 K. The red curve in (e) is obtained from fitting the modified Kittel equation. The red lines in (f) provide a guide to the eye of the line broadening at low temperatures.}
    \label{fig:SRT+anis}
\end{figure*}

Figure \ref{fig:SRT+anis}a displays a graph of the areal anisotropy density $K_{eff}\cdot t_{Fe}$, obtained from SQUID-VSM magnetometry measurements at room temperature, as a function of Fe thickness. The effective anisotropy $K_{eff}$ for each point was extracted from the magnetization work area difference $W_{IP}-W_{OOP} = 2K_{eff}$ \cite{CullityGraham} between M(H)-curves for each $t_{Fe}$, measured under in-plane and out-of-plane magnetic field application (Supplementary Figure S3a-e). The observed linear trend is consistent with equation (\ref{Gradmann}). Clear ferromagnetic switching and estimated $M_s$-values within 1700 $\pm$ 170 kA/m indicate robust ferromagnetism down to sub-nm thickness. This is consistent with the structural continuity of the 0.8 nm Fe film, evidenced in the previous section, avoiding superparamagnetic behavior due to island-like or granular morphology. The SRT is found as the value of thickness $t_{Fe}$ for which $K_{eff}$ approaches zero, i.e., where the equilibrium magnetization orientation transitions from in-plane to out-of-plane. It is intuitively clear that any modification to the Fe magnetic state by the interfacial value $K_{s,Fe/SrTiO_3}$ will be most impactful near the SRT. We estimate $t_{SRT} = $ 0.7 nm from the intersection of the linear trend with the horizontal \textit{t}-axis. The $K_{eff}\cdot t$-intercept represents the total interface anisotropy $K_{int} =K_{s,Fe/SrTiO_3} + K_{s,top}$ and the extrapolated value is 1.4 $\pm$ 0.1 mJ/m$^2$. We can estimate the contribution from the Fe/SrTiO$_3$ interface as:

\begin{equation}
    K_{s,Fe/SrTiO_3} = K_{int}-K_{s,top} 
\end{equation}

Although the top interface is chemically inhomogeneous, we can use the value of $K_{s,top} \approx K_{Fe/Ag}$ as an upper boundary, since the Fe/Ag interface anisotropy is larger than for Fe/Au or Fe/Al \cite{HeinrichCochran, FirstPrinciplesMetallicCap}. Experimental literature values of the Fe/Ag anisotropy vary between 0.64 and 0.8 mJ/m$^2$ \cite{Fe_Ag_Anisotropies, FeAgSRTWedges}. Hence, we estimate  $K_{s,Fe/SrTiO_3} \approx$ 0.6-0.8 mJ/m$^2$. Similar values have been reported in other ab-initio works \cite{STOFeMCADFT, STOFeVCMADFT}. As a first-principles verification of our anisotropy characterization, density functional theory (DFT) calculations were performed for idealized heterostructures of Ag/Fe/SrTiO$_3$ consisting of well-defined atomic monolayers. Their $K_{eff} \cdot t$-values are plotted as green points in Figure \ref{fig:SRT+anis}a. The structure of 3 ML Fe is shown in Figure \ref{fig:SRT+anis}d, with layer-resolved magnetic anisotropy values. It can be seen that the values computed by DFT extrapolate to a larger value than our experimental results. However, the interfacial contribution of the two Fe ML nearest to the Fe/SrTiO$_3$-interface in Figure \ref{fig:SRT+anis}d is quite close to our experimental lower boundary of $K_{s,Fe/SrTiO_3}$. The discrepancy at very low thickness can be explained partially by the inhomogeneity of the Fe top interface in our samples, reducing the Fe/Ag interface contribution. 

\vspace{10 pt}

Using our experimental identification of the SRT, we choose to focus on the sample with $t_{Fe}$ = 0.8 nm, which exhibits a small negative (hence in-plane favored) $K_{eff}$-value. Figure \ref{fig:SRT+anis}b and c show the $M(H)$-curves and corresponding magnetic anisotropy energy $2K_{eff}$ (green shaded area) at T = 300 K and 4.2 K, respectively. It is clear that the anisotropy energy further decreases in the cryogenic regime, evidenced by the reduced area between the loops. While the evolution of the effective, global anisotropy $K_{eff}$ is clear, it is important to note that this value varies on a local scale, closely linked to the film thickness variations and defect profile observed in Figure \ref{fig:structure}b. The film consists of Fe regions or `patches' with varying thickness and anisotropy, which average to a near-zero $K_{eff}$ value over the film volume close to the SRT. This physical picture is strengthened by corresponding ferromagnetic resonance (FMR) measurements, shown in Figure \ref{fig:SRT+anis}e and f. The resonance line in Figure \ref{fig:SRT+anis}e was fitted by the modified Kittel equation, yielding an estimated interface anisotropy of $K_{int}$ = 1.3 $\pm$ 0.2 mJ/m$^2$ (Supplementary Figure S3f). The magnetic field during FMR was applied in the film plane, along the Fe $<$100$>$ direction. When cooled to 5 K (Figure \ref{fig:SRT+anis}f), the resonance line exhibits strong inhomogeneous broadening. This is attributed to sharp local magnetic anisotropy variations, linked to the nonuniform Fe thickness and microstructure. At low temperatures, thermal activation over local anisotropy barriers is reduced, and the resulting inhomogeneous anisotropy field enhances the damping contribution. Given the insights from our structural and magnetic analysis, we identify the $t_{Fe}$ = 0.8 nm sample as an inhomogeneous, yet continuous ferromagnetic system near the SRT instability, well-suited for the investigation of VCM in the cryogenic regime, where its $K_{eff}$ value is sufficiently low to achieve substantial impact of voltage-induced modifications.

\subsection{Magnetoelectric characterization by magnetotransport measurements} \label{transport}

\begin{figure*}[h!]
    \centering
    \includegraphics[width=7.2 in]{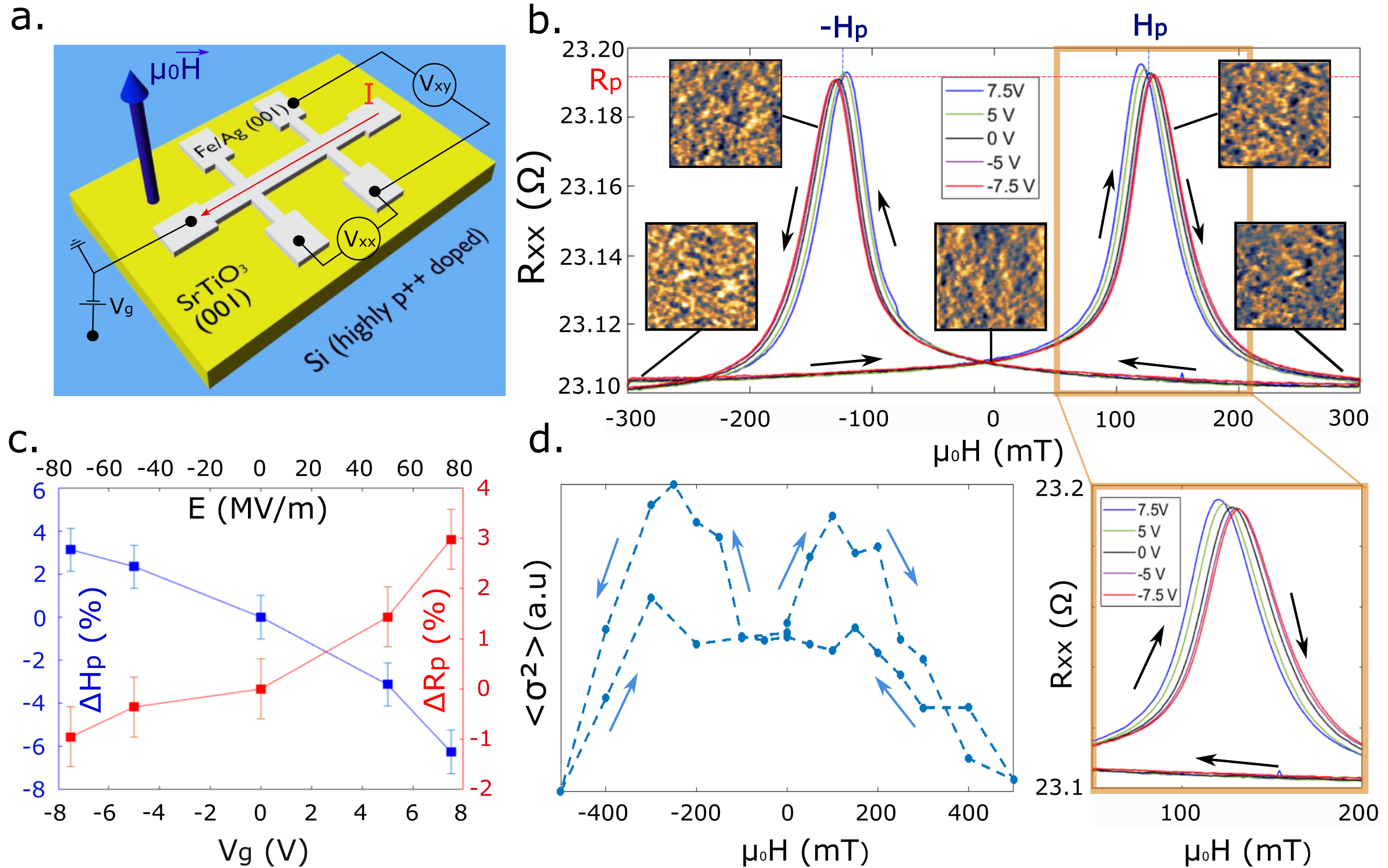}
    \caption{Magnetotransport measurements of the Fe/Ag system at $T$ = 6 K, with $\mu_0$H applied perpendicular to the current plane. (a) Measurement configuration for $R_{xx}$ and $R_{xy}$, with gate voltage applied between the Fe/Ag bridge (ground) and silicon substrate (signal). (b) $R_{xx}(H)$ measurement, with MFM-images (obtained at $T$ = 4 K) in the insets, visualizing the corresponding magnetic microstructure. The extracted window shows the resistance peaks in more detail. (c) $H_p(V_G)$ and $R_{xx}(V_G;H_p)$-dependences, expressed as a percentage of the $H_p(0)$ and $R_{xx}(0, H_p)$- $R_{xx}(0, 0)$ values, respectively. The approximate linear trend is consistent with DFT-predictions. (d) Extracted average variance $<\sigma^2>$ of the MFM images at each magnetic field, correlated to the degree of overall magnetic misalignment. }. 
    \label{fig:transport}
\end{figure*}

A sensitive and application-relevant probe to detect and quantify VCM is through magnetotransport measurement. Magnetotransport properties and corresponding back gate-induced modifications were assessed by electrical characterization in a Liquid Helium cryostat. Transport structures (Hall bars, dimensions 2.3 mm x 130 $\mu$m), schematically depicted in Figure \ref{fig:transport}a, were fabricated from the $t_{Fe} = 0.8$ nm thin film samples by Xe-ion milling (Experimental Methods). The longitudinal ($R_{xx}$) and transverse ($R_{xy}$) magnetoresistance were measured at a temperature of $T$ = 6 K, well within the expected high-permittivity temperature regime of SrTiO$_3$ \cite{khalil2026cryogenicpiezoelectriceffectsfilm}, under an out-of-plane magnetic field and bias current $I_{bias} = 500$ $\mu A$. This configuration allows to investigate both the conventional magnetoresistance and the Hall response, while it competes with the small native in-plane magnetic anisotropy $K_{eff}$ of the Fe film. A back-gate voltage was applied between the Fe/Ag transport bridge (reference ground) and the conductive silicon substrate (voltage signal), as shown in Figure \ref{fig:transport}a. No leakage currents in excess of 100 nA were detected during measurement.

\vspace{10 pt}

The longitudinal magnetoresistance measurement $R_{xx}(H)$ (Figure \ref{fig:transport}b) shows a very prominent field dependence, with pronounced resistance peaks near magnetic fields $\pm H_p$. The effect of back-gate voltage on the $R_{xx}(H)$-response is clearly noticable, inducing substantial relative shifts in the $H_p$-values, as well as slight modifications of the $R_p = R_{xx}(H_p)$ values. This is shown in Figure \ref{fig:transport}c, expressed as a percentage of the $H_p$ and $[R_{xx}(0, H_p)$- $R_{xx}(0, 0)]$-values at 0 V gate. The transverse magnetoresistance $R_{xy}(H)$, corresponding to the Hall contribution, exhibits strictly linear behavior as a function of $\mu_0H$ (Supplementary Figure S5). The absence of an Anomalous Hall contribution $R_{AHE} \propto M_z$ in $R_{xy}(H)$, combined with sizeable longitudinal response, implies a dominant contribution of spin-dependent scattering at the Fe/Ag interface \cite{SpinDependentScattering, LevyZhang96, Parkin_spin_dependent_scattering, StilesSDS96}. Such mechanism does not generate a transverse component, while the longitudinal component generally scales with the degree of spin misalignment or `disorder'. \cite{MILLS19661805}:
\begin{equation}
    \rho_{xx} = \rho_{xx,0} + \rho_{xx, \uparrow\downarrow}(H)
    \label{DeGennes}
\end{equation}

While the spin-dependent term $ \rho_{xx, \uparrow\downarrow}$ is intricately linked to the magnetic microstructure and scattering profile during reversal, the $R_{xx}(H)$ measurement provides a qualitative probe of the Fe ferromagnetic domain configuration. To substantiate the qualitative validity of equation (\ref{DeGennes}) for our system, Magnetic Force Microscopy (MFM) measurements (15 x 15 $\mu$m$^2$ scan size) were performed on a similar sample, deposited under identical growth conditions. These are shown as insets in Figure \ref{fig:transport}b, permitting visualization of the ferromagnetic microstructure during the magnetic field sweep. Images for the full magnetic field dependence can be found in Supplementary Figure S8. Near the $H_p$-values,  the domain irregularity and overall magnetic inhomogeneity is enhanced. As a quantitative representation for this rather subtle visual change, Figure \ref{fig:transport}d shows the mean phase variance value $<\sigma^2>$ of the MFM images, which provides a useful approximate estimate of the overall degree of spin misalignment in the system. The result qualitatively reproduces the $R_{xx}(H)$-dependence (with slightly higher $H_p$-values). A more in-depth discussion on the procedure to obtain the $<\sigma^2(H)>$ plot can be found in the Supplementary Information section 4.2 (Figure S5). 

\vspace{10 pt}

Irrespective of the detailed spin-dependent scattering contributions leading to the $R_{xx}(H)$-dependence, the longitudinal magnetoresistance measurements reveal a clear effect of back-gate voltage on the Fe ferromagnetic configuration. For positive voltages, the $H_p$-value shifts to lower absolute values, while for negative voltages it increases (Figure \ref{fig:transport}b,c). Furthermore, the longitudinal magnetoresistance peak $R_p$ shifts to higher values for increasing positive voltages. The relative shifts in these quantities are expressed as a percentage of the peak field $H_p$ and magnetoresistance $\Delta R_{xx} = R_{xx}(H_p) - R_{xx}(0)$ at 0 V gate. First-principles calculations (Supplementary Figure S4) predict that positive electric fields reduce the interfacial anisotropy energy component $|K_{int}|$, favoring more in-plane magnetization in the overall system. Conversely, negative electric fields enhance perpendicular magnetic anisotropy. This is consistent with our observations. Due to the complexity of the magnetic domain reversal and scattering mechanism, it is not straightforward to quantify the VCM effect by a single coefficient $\xi$. However, we can assume that the reversal occurs by domain nucleation and pinning-limited domain-wall propagation, which is very plausible for imperfect ultrathin films, and will be supported by micromagnetic simulations in the next section. In this case we can link the field $H_p$ to an effective domain wall pinning barrier contribution $\Delta_{DW} = \eta K_{Fe/SrTiO_3}$. The $\eta$ value relies on a linear approximation, and depends on the degree of anisotropy variation within the sample and the pinning disorder lengthscale (discussed in Supplementary Information section 4 and Figure S6). 

\begin{equation}
    \xi = \frac{\delta K_{Fe/SrTiO_3}}{\delta E} \approx \frac{1}\eta\frac{\mu_0\delta H_pM_st_{Fe}}{\delta E} 
\end{equation} 

where we estimate $\eta \approx \mathcal{O}(1)$. Filling in the values $\mu_0\delta  H_p$ = -12 mT, $\delta E$ = 0.15 V/nm, $M_s = 1700$ kA/m and $t_{Fe} = 0.8$ nm, we obtain a value of $\xi$ in the order of -100 fJ/Vm, the same order of magnitude as first-principles predictions in literature \cite{STOFeVCMADFT} and our own calculations (Supplementary Figure S4). Note that this value relies on the assumption of domain wall pinning by local anisotropy variations, and a direct relation between the required Zeeman energy and the average domain pinning barrier, determined by the interface anisotropy. Such first-order quantification does not provide an exact value of the VCM coefficient, and should not be interpreted in the strict sense. However, it emphasizes the fact that the observed effect shows overall consistency with first-principle modelling, both in polarity and in order of magnitude. Most importantly, since we are able to position the magnetic system close to the SRT at low temperatures, the voltage-induced $\delta K_{eff} \approx \xi \delta E$ of this energetic scale should be able to strongly modify the equlibrium magnetic state, using a CMOS-friendly and scalable approach at low gate voltages. We explore this systematically in the next sections.

\subsection{Spatially-resolved microscopic magnetoelectric characterization} \label{spatially_res}

\subsubsection{Micromagnetic simulations}

The impact of gate voltage on the Fe/Ag magnetotransport behavior, as characterized and quantified in the previous section, indicates a substantial and polarity-consistent VCM effect. Going one step further, an investigation of VCM focused on the spatially-resolved magnetic microstructure and physical observables directly proportional to the absolute magnetization, would strengthen the physical understanding of the magnetic behavior at the microscopic scale. As such, it provides a more straightforward verification of the gate voltage impact on the ferromagnetic state, while clarifying the link between the magnetic microstructure and the indirect probe of spin-dependent scattering magnetoresistance. 

\begin{figure*} [h]
    \centering
    \includegraphics[width=6.5 in]{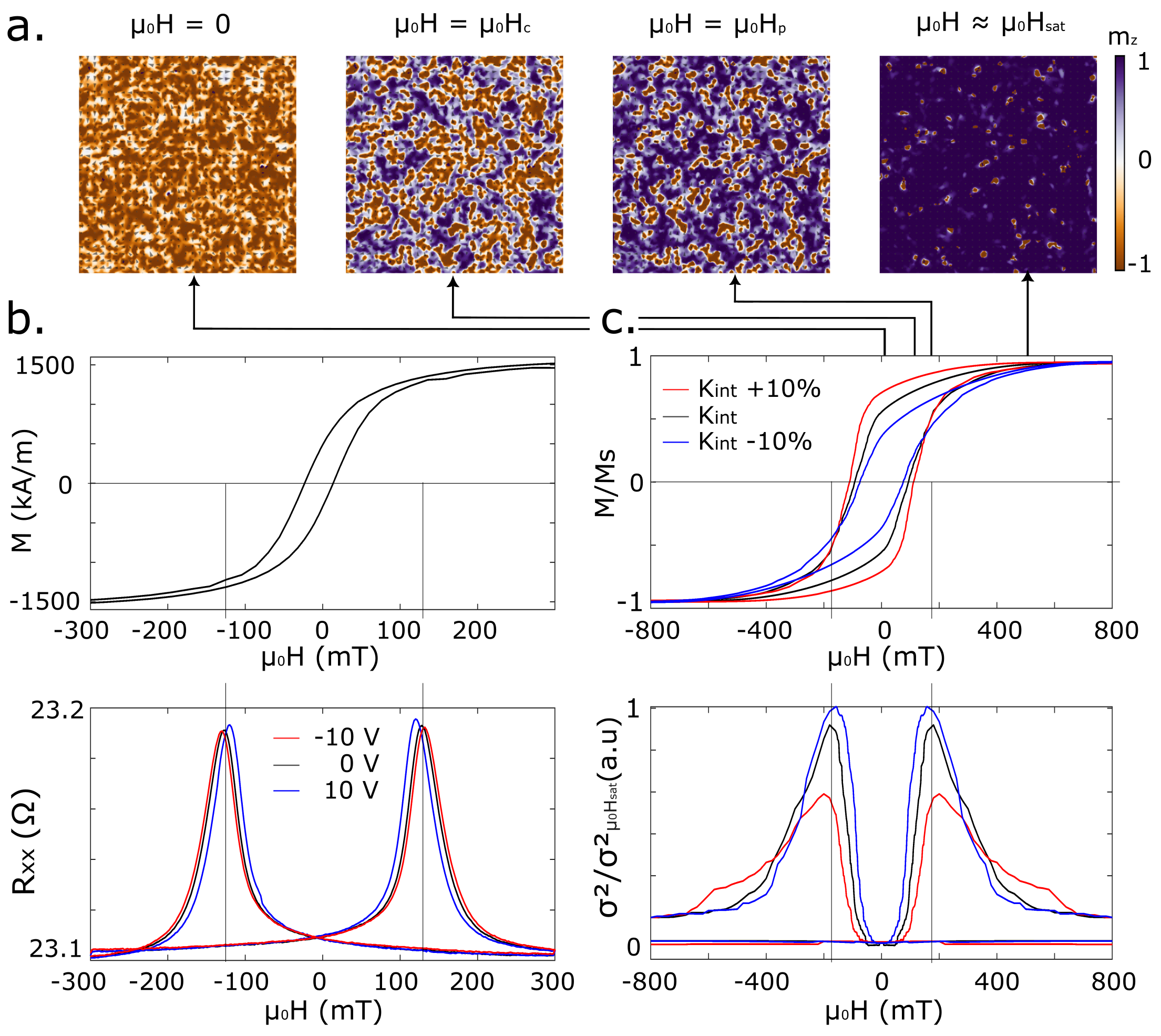}
    \caption{(a) Micromagnetic magnetization maps of the 0.8 nm Fe slab at different magnetic fields during magnetization reversal. Simulation area is 1 x 1 $\mu$m$^2$. (b) Experimentally obtained $M(H)$ (top) and $R_{xx}(H)$ (bottom) loops for magnetic field applied perpendicular to the Fe plane, at back-gate voltage -10 V (red), 0 V (black) and 10 V (blue). (c) Macroscopic normalized  $M(H)$ (top) and $\sigma^2(H)$ (bottom) loops, extracted from the micromagnetic simulations, with positive (red), zero (black) and negative  (blue) $\Delta K_{int}$. The links to the magnetization maps at different field are indicated by arrows.}
    \label{fig:micromag}
\end{figure*}

As a first step, micromagnetic simulations were performed on a 2-dimensional Fe slab with locally varying uniaxial anisotropy $K_u(x,y) = \frac{K_{int}}{t_{Fe}(x,y)}$. A randomized mask of local $t_{Fe}(x,y)$ variations was implemented, on a scale consistent with the Fe TEM and AFM characterization, to approximate the thickness- and compositional inhomogeneity of the Fe film. More details on the micromagnetic parameters and numerical implementation can be found in the Experimental Methods. Figure \ref{fig:micromag}a provides insight in the magnetic microstructure during magnetization reversal in out-of-plane magnetic field, after magnetic saturation at negative magnetic field. Regions with negative and near-zero $K_u$-value (thicker, in-plane favored) gradually align in the film plane as the magnetic field approaches zero. Regions with positive $K_u$-value (thinner, out-of-plane favored) stay oppositely magnetized. As the field increases to the coercive field, domains magnetized along the applied field nucleate preferentially from the low $K_{eff}$ regions, and propagate by reversing the spin orientation of neighboring regions. During this process, the propagating domain boundaries experience regions of high positive $K_u$, as well as local defects like surface terrace edges or crystallite boundaries. The interaction with these locations in the variable anisotropy landscape causes \textit{domain wall pinning} in the field regime between domain nucleation and full saturation of the ferromagnetic system. Moreover, pinned domain walls tend to `buckle' around local pinning sites, causing deformation of the domains into a more irregular shape \cite{KIRILYUK199745}. This phenomenon causes enhanced overall spin disorder, leading to an increase in the magnetoresistance well beyond the nucleation field. When the magnetic field reaches the value $H_p$, the driving force of domain wall motion is large enough to overcome most pinning sites, reducing the global spin-disorder and causing the domain morphology to become more uniform. 

\vspace{10 pt}
The micromagnetic model is validated by comparison of the experimental field-dependences M$_z(H)$ and $R_{xx}(H)$ (Figure \ref{fig:micromag}b) to the ones extracted from the simulation (Figure \ref{fig:micromag}c). The M$_z(H)$-dependence is extracted by a thickness-weighted sum of the $m_z$-components of each simulation cell, while the normalized variance $\frac{\sigma^2(H)}{\sigma^2(H_{sat})}$ is computed as a representative quantity for spin disorder, in accordance with equation (\ref{DeGennes}). Note that we compute the normalized variance only for the regions of thickness below the SRT value ($K_{eff} >$  0), since these are the dominant source of spin misalignment during magnetization reversal. This assumption is required to reproduce the observed difference $H_c < H_p$. The simulation results follow the qualitative behaviour of the experimental data, and reproduce electric field-dependent modifications of $H_p$ and $R_{xx}(H_p)$ for $\Delta K_{int}(E)$ of $\pm 10\%$ (Figure \ref{fig:micromag}c), consistent with the DFT calculations in electric field polarity. It can be noted that the changes in $R_{xx}(H_p)$-values are more drastic than the experimental observations. This is attributed to the complexity of the scattering process ; while the variance gives an insightful qualitative probe of spin-disorder, it can not reproduce the exact values of magnetoresistance. Aside from an intuitive understanding, the micromagnetic results provide a microscopic argument for the electric field impact on the magnetotransport and the VCM quantification. A detailed discussion can be found in the Supplementary Information section 4, but the general argumentation can be stated as follows: since negative gate voltage induces a more positive $K_{int}$ value and hence enhanced perpendicular anisotropy, the anisotropy contrast will increase, resulting in `harder' pinning regions. This leads to more effective pinning of propagating domains and a positive shift in $H_p$. Conversely, a positive voltage reduces the anisotropy contrast, facilitating domain wall motion and reducing the effective pinning strength.
\vspace{10 pt}

\subsubsection{Magneto-Optic characterization}

\begin{figure*} [h!]
    \centering
    \includegraphics[width=7 in]{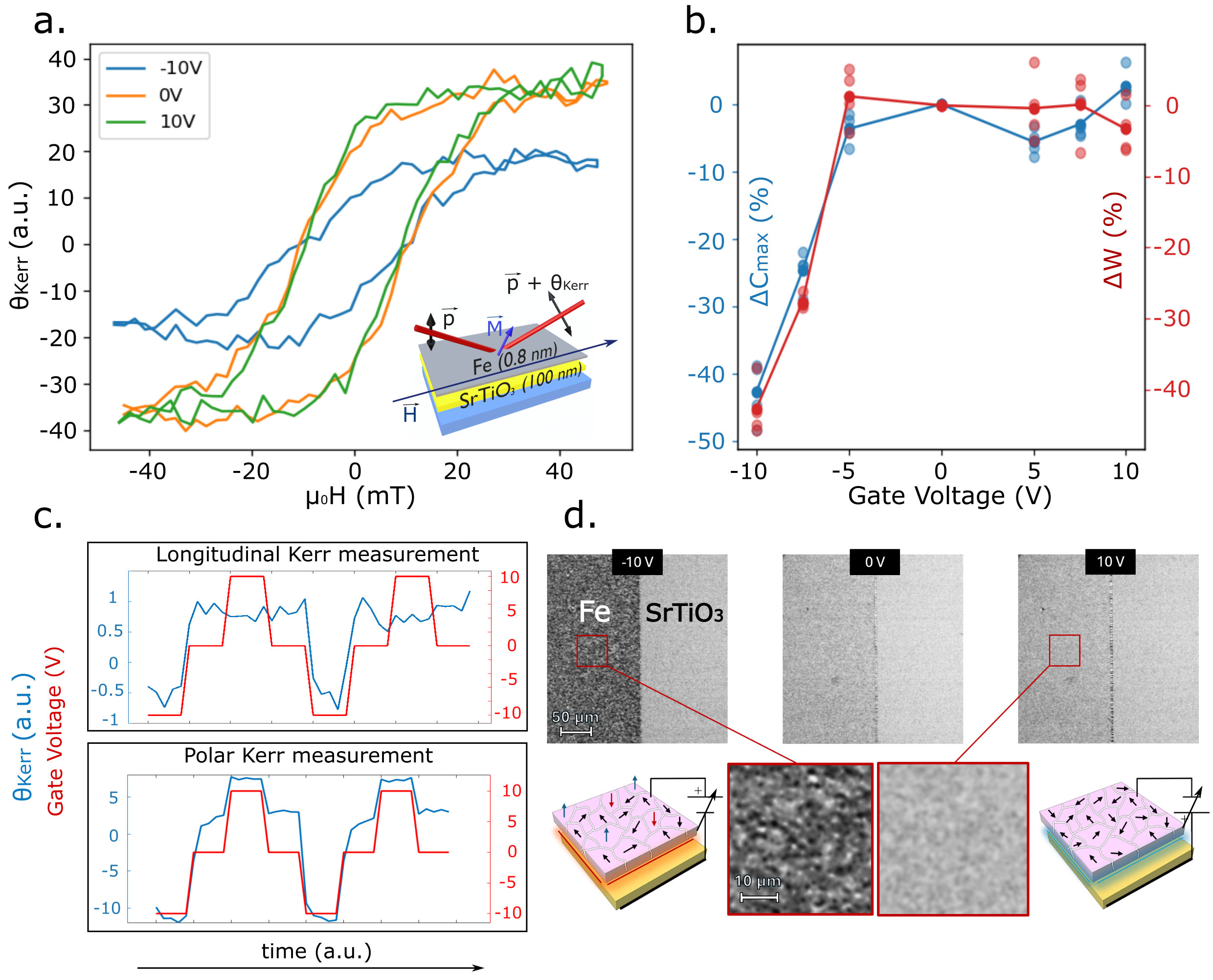}
    \caption{MOKE measurements of the $t_{Fe}$ = 0.8 nm sample under voltage gating. (a) Longitudinal MOKE measurements, showing the in-plane magnetization component along the applied magnetic field direction (Fe$<$100$>$) at $T$ = 5 K. Negative gate voltage reduces the observed amplitude contrast $C_{max} = \theta_{Kerr,L}$(50 mT) -  $\theta_{Kerr,L}$(-50 mT) and hysteretic area of the $\theta_{Kerr}(H)$ loop, indicating a reduction in the in-plane magnetization component. (b) Gate voltage dependence of the relative change in Kerr amplitude contrast $\Delta C_{max}$ and hysteretic area $\Delta W$. (c) Evolution of Longitudinal and Polar Kerr signals in the Fe region while applying back-gate voltage pulses. (d) Polar Kerr measurements, reflecting the out-of-plane magnetization component at zero magnetic field ($T$ = 30 K). The left hand side of the microscope image is the Fe patterned film region, while the right hand side is the bare SrTiO$_3$ film surface. An enhanced polar contrast is observed for negative gate voltage, consistent with the partial reorientation of magnetic domains towards the out-of-plane direction. Enlarged images reveal a strong contrast enhancement at -10 V. }
    \label{fig:MOKE}
\end{figure*}

While our microscopic understanding of the Fe magnetic configuration and the VCM effect exhibits a good qualitative match with the magnetotransport measurements, a \textit{direct} experimental demonstration of the Fe magnetization under varying gate voltage application would provide convincing visual evidence of voltage-controlled magnetism. To this end, Magneto-Optic Kerr-Effect (MOKE) measurements were performed at cryogenic temperatures in a closed-cycle cryostat. The Kerr rotation of the light beam polarization $\theta_{Kerr}$, measured after reflection from the Fe film surface, provides a directly proportional measure of the overall magnetization along the selected measurement direction : $\theta_{Kerr} \propto M_{Fe}$. For the first set of measurements, an in-plane magnetic field $\mu_0 H_{IP}$ was varied along the Fe$<$100$>$ direction, transverse to the transport bridge. The \textit{longitudinal} Kerr rotation $\theta_{Kerr,L}$, proportional to the magnetization along the  $\mu_0 H_{IP}$ direction, was recorded for different back-gate voltages. The measurement was done at $T$ = 5 K and in a magnetic field range of $\mu_0 H_{IP}$ = $\pm$ 50 mT (Figure \ref{fig:MOKE}a). The longitudinal loop obtained at 0 V gate (orange) has a coercive field and shape consistent with the M-H loop obtained by SQUID-VSM (blue curve in Figure \ref{Magnetometry}c.). Since full saturation is never reached within the applied magnetic field range, it represents a \textit{minor} $M(H)$ hysteresis loop, directly proportional to the Fe magnetization along the in-plane field direction. 

\vspace{10 pt }

The effect of negative gate voltage (-10 V) on the LMOKE hysteresis loop is clearly reducing the ability to magnetize along the field axis. Additional loops for intermediate voltages are shown in the Supplementary Figure S7. As a first-order quantification, the average values of LMOKE contrast between maximal field values $C_{max} =$ $\theta_{Kerr,L}$(50 mT) -  $\theta_{Kerr,L}$(-50 mT) and the loop area $W = \int\theta_{Kerr}(H)dH$, are shown in Figure \ref{fig:MOKE}b. For negative voltages, both values become strongly reduced compared to the 0 V loop. For positive gate voltages, no clear gate effect is observed . Since the magnetic field is applied in the Fe film plane, this implies a \textit{reduced in-plane} tendency of the magnetization for negative voltage, consistent with the \textit{enhanced out-of-plane} tendency observed in the magnetotransport measurements under out-of-plane field and first-principles calculations. Hence, our voltage gating results are consistent with the physical description of polarity-dependent interfacial anisotropy shifts for in-plane (MOKE), as well as out-of-plane (transport) magnetic field  orientations. The reason for the apparent absence of a gate-effect at positive voltages in Figure \ref{fig:MOKE}b could suggest a non-linear dependence of magnetic anisotropy on gate voltage. However, since in-plane magnetization is already favored at 0 V ($K_{eff}$ $<$ 0), the impact of `strengthening the easy axis' along the field direction is less apparent than inducing an anisotropy term that competes with the applied magnetic field. 

\vspace{10 pt}

Based on the results obtained so far, we have established that negative gate voltage enhances the perpendicular magnetic anisotropy component that competes with the pristine, slightly in-plane favored effective anisotropy $K_{eff}(0)$. The most direct approach to demonstrate a practically useful VCM effect, would be to measure  the magnetic contrast in absence of a magnetic field (`field-free switching'). We implement this measurement configuration by first magnetizing the sample in-plane at $T$ = 30 K and $V_G$ = 0 V, and subsequently applying $V_G$ = -10 V, 0 V and 10 V back-gate pulses. Following the measurement protocol described in \cite{McCord_2015}, we measure the averaged \textit{longitudinal} (LMOKE) and \textit{polar} Kerr (PMOKE) signal evolution, sensitive to the \textit{in-plane} and \textit{out-of-plane} magnetization component, respectively. These measurements allow to correlate the magnetic response with the back-gate excitation. Figure \ref{fig:MOKE}c shows the evolution of LMOKE and PMOKE signal under gate application, relative to the 0 V baseline. The gate voltage sequence and magnetic response are clearly correlated, with an apparent correlation between PMOKE and LMOKE signal. When negative gate voltage is applied, perpendicular anisotropy is promoted. Consequently, the LMOKE signal reduces, while the absolute value of the PMOKE signal is enhanced. For positive voltage, the LMOKE signal does not appreciably change, while the PMOKE signal is slightly reduced. This is caused by forcing remaining out-of-plane fractions more in-plane. While these macroscopic observations are consistent with our understanding of the VCM mechanism, it would be highly desirable to image the magnetic domain structure during gate voltage application. 

\vspace{10 pt}
Since the PMOKE signal is 10 times more intense than the LMOKE signal, and hence very sensitive to the out-of-plane magnetization component, we choose to focus on the PMOKE-measurement for a spatially-resolved measurement of the VCM effect, shown in Figure \ref{fig:MOKE}d. Upon application of a -10 V gate voltage, the PMOKE contrast increases significantly, linked to the formation of out-of-plane oriented or canted magnetic domains. The magnetic microstructure is inhomogeneous and appears qualitatively similar to the domain structure imaged by MFM (Supplementary Figure S8 and S9), but with lower spatial resolution and at a larger imaging scale. For 0 V and 10 V, no substantial difference is observed between the Fe and SrTiO$_3$ signal, suggesting predominantly in-plane magnetized magnetic domains for both cases. This is conceptually illustrated in the schematic depictions adjacent to the PMOKE images. The clear correlation of the longitudinal and polar Kerr signal variations to the back-gate pulse sequence, as well as the visual demonstration of enhanced out-of-plane domain nucleation at negative voltage, shows the ability to electrically modify or `write' the ferromagnetic state in a reversible manner. Since no remanent effect of the gate voltage was observed, this writing is volatile. While this implies that the VCM effect cannot be used to write a persistent magnetic state, the temporary gate-induced modifications of the magnetic configuration could nevertheless be highly impactful for cryogenic device applications, as we will point out in the discussion.

\section{Discussion and Conclusion} 

In this work, we demonstrated and characterized cryogenic Voltage Control of Magnetism (cryo-VCM) in a thin film SrTiO$_3$/Fe/Ag heterostructure, integrated on a silicon substrate. The epitaxial deposition resulted in a heterostructure with high crystalline quality and sharp Fe/SrTiO$_3$ interface. By systematic optimization of the Fe thickness, we selected a sample that was energetically very close to the Spin Reorientation Transition (SRT) and hence most sensitive to changes in the magnetic anisotropy at the Fe/SrTiO$_3$ interface. The magnetic configuration of the Fe film in this sample was inhomogeneous, as demonstrated by Magnetic Force Microscopy, largely due to the ultrathin and locally varying thickness. From magnetotransport characterization, we found that back-gate voltage application substantially affected the magnetization reversal process, which we attributed to electric field-induced changes in the magnetic anisotropy at the Fe/SrTiO$_3$ interface, consistent with our first-principles calculations. From these measurements, we were able to obtain a quantitative experimental estimate of the VCM coefficient, competitive with state-of-the-art implementations of electronic VCM \cite{VCMAreviewNozakiMiwa}. By developing a micromagnetic model that reasonably reproduced the VSM magnetometry, magnetotransport and MFM results, we established a physical understanding of the probable manifestation of the VCM impact on the microscopic magnetic configuration. Magneto-optical measurements provided a direct insight into the gate voltage impact on the global magnetization ($M(H)$) process, as well as spatially-resolved imaging of the inhomogenous magnetic domain contrast under gate voltage application. We demonstrated clear and reversible modifications of the out-of-plane magnetization component under electric field application. Our experimental findings consistently point out the significant potential of the SrTiO$_3$/Fe heterostructure on conductive Si substrate as a CMOS-friendly gate-controlled nanomagnetic material system for cryogenic electronics. 

\vspace{10 pt}

\begin{figure*} [h]
    \centering
    \includegraphics[width=6.5 in]{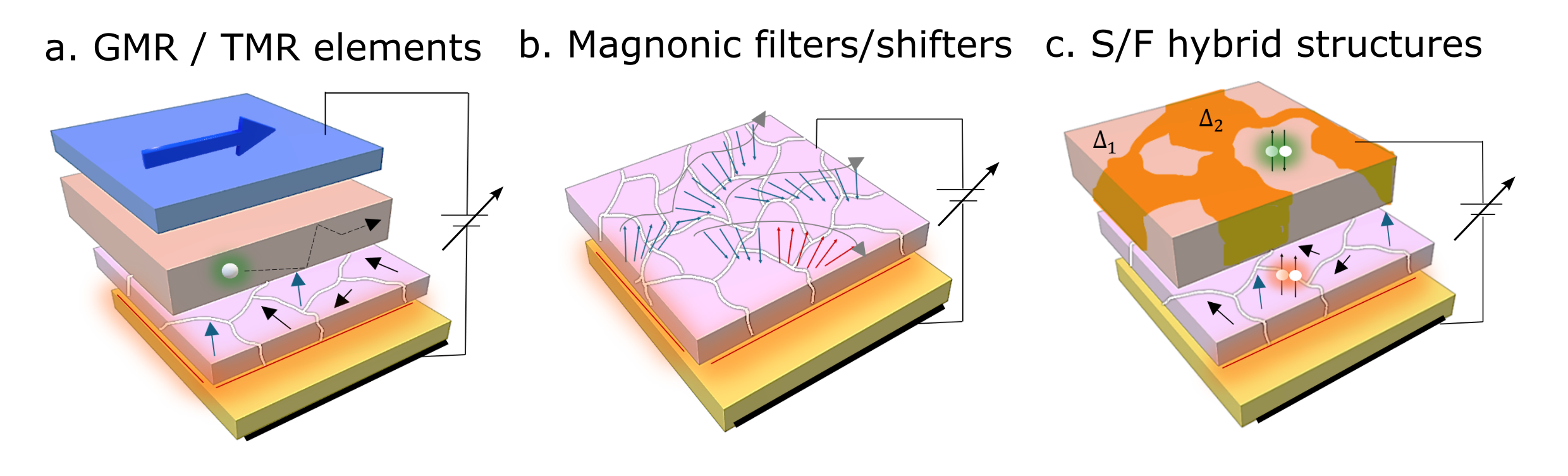}
    \caption{(a) Low-power GMR elements with electrostatically-controlled 'free layer' (b) Magnonic material for electric control of spin-wave propagation (c) Superconductor-ferromagnetic hybrid material, hosting proximity effects that couple the superconducting order parameter to the ferromagnetic configuration. Electrical control of the magnetic domain structure allows gate-tunability of supercurrent properties.}
    \label{fig:Device implementations}
\end{figure*}

The ability to control magnetic domain configurations in our system by electric field at cryogenic temperatures forms a highly attractive physical property in the context of low-power spintronics and integration with superconducting electronics. In particular, we highlight three promising application avenues (Figure \ref{fig:Device implementations}), for which we believe this system in particular holds strong potential. \textit{Giant- and tunnel magnetoresistance} (GMR/TMR) junctions are extensively used in random access and neuromorphic computing technology. Electric field-induced lowering of the magnetic anisotropy barrier reduces the switching energy of the ultrathin 'free' magnetic layer, while the magnetic inhomogeneities in this layer could be exploited to enhance interfacial exchange scattering contributions to the device resistance. At low temperatures, reduced thermal noise further improves switching fidelity and temporal precision. Additionally, downscaling of the lateral dimensions reduces the geometric capacitance of the junction element, leading to faster switching times. Another relevant domain would be \textit{magnonics}, where voltage-tunable anisotropy allows the dynamic reconfiguration of spin-wave dispersion and band gaps through shifts in the effective anisotropy field, enabling low-dissipation frequency tunability and phase control \cite{VcontrolledMagnonics}. When combined with carefully engineered device structuring, the presence of gate-tunable inhomogeneous magnetization can be used to confine, filter, phase-shift or interfere spin waves and induce non-reciprocal response. The FMR results of Figure \ref{fig:SRT+anis}e,f indicate that, even for ultrathin dimensions, the Fe film still exhibits a sizable radio frequency response. Finally, a highly interesting application perspective would be the integration into a \textit{superconducting/ferromagnetic hybrid} material stack \cite{Superspintron, Eschrig_2015}, where the inhomogeneous magnetization profile can be translated into superconducting gap variations, inducing a spatially varying amplitude and phase of the condensate. In the ferromagnetic component, superconducting proximitization can induce so-called \textit{spin-triplet} supercurrents, which remain coherent over substantial lengthscales and whose phase depends on the local magnetic environment. A gate-controlled magnetic source of superconducting gap variations and triplet supercurrents opens serious potential for scalable, gate-controlled and phase-tunable ($\varphi_0$) Josephson junctions, which could dramatically reduce design complexity and enhance implementation flexibility of quantum- and classical superconducting circuitry.

\vspace{10 pt}

\pagebreak
\section{Experimental Methods}
\threesubsection{Thin film deposition}

Sr was evaporated from a dual-filament Knudsen effusion cell while Ti was generated by an electron beam evaporator. The flux of both metallic molecular beams was calibrated in situ using a quartz crystal microbalance (QCM). The Si (001) wafers were cleaned for 90 s in a 2\% HF solution before they were introduced to the ultrahigh vacuum (UHV) growth chamber. Sr-assisted native oxide desorption further prepared the atomically clean surface required for direct epitaxy. Before starting the STO growth, 1/2 monolayer of Sr was first grown, acting as an oxidation barrier between Si and STO. Direct epitaxy of the first 3 nm of STO was performed in molecular oxygen at 350°C. After this, the growth was paused to switch to atomic oxygen (600 W plasma power) in the growth chamber and to increase the substrate temperature to 550°C. Under these conditions, the remaining STO epitaxy was completed at a growth rate of approximately 1 nm/min (1.1E-6 Torr). For the STO layer discussed in this work, the Sr cell temperature was gradually increased by 11°C to counteract source oxidation. In the middle and at the end of the growth, the growth was paused for an in situ annealing in oxygen atmosphere to 800°C to improve crystallinity, surface roughness, and oxygen stoichiometry. Afterwards, the sample was cooled down at 10 °C/min (~2E-6 Torr). The average cationic stoichiometry through the layer obtained from Rutherford backscattering spectroscopy (RBS) is Sr/Ti = 0.94 +/- 0.03. After deposition of the SrTiO$_3$ thin films, samples were taken out of the UHV environment. The SrTiO$_3$ surface was then treated for heteroepitaxial growth : the samples were immersed in deionized water for 20 minuted under mild sonication, followed by a 30 second dip in ammonium fluoride buffered HF solution (NH$_4$F:HF, 87.5:12.5 from Honeywell). This step removes SrO-islands resulting from strontium secretion after MBE-growth. After rinsing again with deionized water, the samples were loaded in a tube furnace oven and heated for 60 minutes at 850 degrees Celsius under O$_2$ flow (flow rate around 200 sccm), to achieve a flat and atomically terraced surface structure and improve crystallinity. The O$_2$ flow prevents oxygen vacancy formation during high-temperature treatment. Samples were then loaded in a Riber MBE chamber (base pressure 6 E-11 Torr) for deposition of the metallic ultrathin film stack. Prior to deposition, a brief (15 min) outgassing at 300 degrees Celsius was performed to remove residual water and hydrocarbon contaminants. The sample was cooled down to room temperature, after which the ultrathin Fe film was deposited by electron gun evaporation, at a growth rate of 0.03 \AA/s. The deposited film was annealed to 250 degrees Celsius, to promote coalescence of the dense Fe islands into a continuous, epitaxial film. After cooling back down to room temperature, the Ag capping layer was deposited by thermal K cell evaporation at 0.24 \AA/s. The final protective Al-capping was deposited at room-temperature, by electron gun evaporation at 0.15 \AA/s.

\threesubsection{Volumetric magnetometry measurements}

SQUID-VSM measurements of the M-H hysteresis curves were performed in a Quantum Design MPMS3 system. The samples were cleaved after deposition, using a diamond-tipped pen, into dimensions below 5 x 5 mm to fit the sample holder. The film area for magnetic volume estimate was based on the measurement of the sample edges, and has an error of $\pm 0.5$ mm. For in-plane magnetic field measurements, the sample was attached by GE-varnish to a quartz paddle holder. For out-of-plane field measurements, the sample was clamped in the center of a plastic straw. Diamagnetic background substraction, originating from the Si-substrate was corrected using a linear fit of the M(H)-curve in the field region between 4 and 5 T. Ferromagnetic resonance measurements were done in a PPMS Dynacool cryostat, using a Keysight Vector Network Analyzer. The sample was pasted on the PCB holder, with the film facing the RF transmission line. The S$_{21}$ transmission coefficient was measured for frequency sweeps from 10 kHz to 20 GHz, while the magnetic field was swept from 300 mT to -300 mT in steps of 0.3 mT 
(3 Oe).

\threesubsection{Density Functional Theory}

The density functional theory calculations were performed using the VASP6.5.1 abinitio package \cite{DFTref1, DFTref2, DFTref3} The Ag/Fe//SrTiO$_3$ (STO) thin film heterostructure is modeled as 3 layers of SrTiO$_3$, terminated on both sides by a TiO surface to avoid remanent dipolar field, followed by 3 layers of Fe, 2 layers of Ag and followed by more than 30 \AA\ of vacuum. Cubic STO, with bulk lattice parameter of 3.905 \AA\ was kept fixed during all calculations. The unit cell of the Fe and Ag layers were rotated by 45 degrees as compared to STO to minimize the in-plane strain in agreement with previous work \cite{STOFeVCMADFT}. PAW pseudo potential were used with the following valence electrons \cite{DFTref4}: Sr 5s, Ti 3d and 4s, O 2p and 2s, Fe 3d and 4s, Ag 4d and 5s. The revised Perdew, Burke, and Ernzerhof functional for solids (PBE$_{sol}$) \cite{DFTref5} is used for all calculations. The Ag/Fe thin film is first relaxed using all symmetry with a 15x15x1 k-point mesh and a plane wave basis energy cutoff of 520 eV until the forces were smaller than 0.001 eV/\AA. After relaxation, the Spin orbit coupling was included with non-collinear magnetism as a function of electric field. During this step, we have approximated STO as a perfect dielectric and neglected atomic relaxation as a function of electric field. Therefore, electric field is obtained by fitting the potential within the STO layer. The quantization axis is varied between the x-, y- and z-direction to determine the magnetocrystalline anisotropy. The MAE is computed as the difference of total SOC contribution to the total energy between the z- and x-direction.

\threesubsection{Transport measurements}
Hall bar structures were patterned by Electron Beam Lithography using an MMA-PMMA resist stack, followed by development in MIBK solution. The developed structures were physically etched into Hall bars using Xe-ion milling (with Ar as neutralizer), after which remaining resist was removed by immersion in warm (60 degrees Celsius) acetone. During the milling process, gas flow controllers were set to 2 sccm for Xe and 2.4 sccm argon, resulting in a pressure equilibrating to 2.9E-4 Torr during the waiting times. Duty cycles of 10 s milling / 40 s waiting were used and He-flow was applied at the backside of the holder, to prevent excessive heating of the sample. The Hall bar samples were wire bonded manually using Au-wire and silver paint onto the sample holder of the Oxford Instruments Heliox He-cryostat. The Hall bar was always grounded to the cryostat ground, while the conductive Si back-gate was held at the gate voltage $V_g$, provided by a Keithley 2400 source meter in voltage mode. Bias current was 500 $\mu$ A for each measurement, provided by a second Keithley 2400 source meter. The magnetic field was always applied perpendicular to the sample surface and swept between 300 mT and -300 mT in steps of 1 mT. 

\threesubsection{MOKE measurements}

The sample was installed in a closed-cycle cryostat (Montana Instruments) and cooled down from RT to 5 K. The sample was identical as the one used for transport measurements, and bonded again with the Fe film as electrical ground contact. A x50 objective was installed inside the cryostat and maintained at room-temperature. MOKE images were obtained with wide-field polarization microscope (with fixed polarizer, rotatable analyzer and compensator). The light source was composed of 4 pairs of leds forming a cross-like light source, allowing for the selection of incident angle and Kerr direction sensitivity (longitudinal – in the sample plane and along the applied magnetic field, transversal – in the sample plane and perpendicular to the applied field, and polar - perpendicular to the sample plane). Each LED is a white light source with main intensity peak around 455 nm (best expected resolution around 200 nm).
Both in-plane (in long. and trans. directions) and Polar Kerr effect were extracted by combination of images captured with different incident angles \cite{Soldatov17, Abhishek26}. The hysteresis loops shown in Fig. \ref{fig:MOKE}a were computed by averaging the Kerr contrast of images captured at each applied field step. To ensure reproducibility, the experiment was repeated ten times for each gate voltage, with voltages ranging from -10 to 10 V. The gate voltage sweep presented in Fig. \ref{fig:MOKE}c was obtained in the remanent state following the application of an IP field of +50 mT. Images were recorded continuously while cycling the gate voltage between -10, 0, and 10 V with a square signal.
To compensate for the parasitic Faraday effect, the Fe layer and the STO substrate were imaged simultaneously. The average Kerr signal was then determined by calculating the difference in contrast change between the exposed Fe region and the STO reference region.

\threesubsection{MFM measurements}

The magnetic domain structure was observed at 4 K using a low-temperature magnetic force microscope (AttoCube AttoDry2100). The voltage was applied to the Fe film and STO substrate using a Keithley 2100 source meter, with the Fe film serving as the positive electrode and grounded. Commercial magnetic probes were driven at a resonance frequency of approximately 81 kHz and lifted about 300 nm above the sample surface. Magnetic contrast was detected in non-contact lift mode through the phase signal, with the phase shift proportional to the normal component of the force exerted on the cantilever, where bright color indicates repulsion and dark color indicates attraction. Magnetic field was applied perpendicular to the sample surface for each measurement.

\threesubsection{Micromagnetic simulations}
Micromagnetic simulations were performed using the open-source software MuMax$^3$ \cite{vansteenkiste_design_2014}, which employs the finite-difference method to solve the Landau–Lifshitz–Gilbert (LLG) equation for time- and space-dependent magnetization dynamics. The Fe film was discretized into $3 \times 3 \times 0.8$ nm$^3$ cells, below the exchange length $l_{ex} \simeq 3.4$ nm. The simulation box consisted of $512 \times 512 \times 1$ cells. Periodic boundary conditions were applied along both IP directions to achieve a demagnetization factor $N_{OOP} \simeq 1$. Interfacial magnetic anisotropy was implemented as a uniaxial anisotropy along the OOP direction with an average intensity $K_{u1} + K_{u2} = K_{int}/t$, where $t = 0.8$ nm is the average film thickness and $K_{int} = 1.3$ mJ/m$^2$ is the experimentally measured interfacial anisotropy. The ratio between the first- and second-order anisotropy constants was set to $K_{u2} \simeq 0.9 K_{u1}$. The exchange constant $A_{ex} = 21$ pJ/m and saturation magnetization $M_s = 1.66$ MA/m were chosen close to bulk values; $M_s$ was set slightly below 1.7 MA/m to ensure the effective anisotropy $K_{eff} = 0$ (Spin Reorientation Transition) at $t = 0.75$ nm. To account for variations in Fe thickness and microstructure, the uniaxial anisotropy $K_u$ was spatially varied. The simulation box was divided into regions of constant $K_u$ with an average diameter of 8 nm. The local anisotropy $K_u^i$ in region $i$ was defined as $K_u^i = K_{int}/t_i$, where $t_i$ represents the local thickness. This thickness follows a Gaussian distribution with a mean of 0.8 nm and a standard deviation of 30\%. The exchange constant was reduced by 50\% at the interfaces between adjacent regions to account for inter-grain decoupling. 
Simulations were initialized with a random magnetization distribution and subsequently relaxed under an external field of -800 mT. The external field was then swept from -800 to +800 mT; at each field step, the system was allowed to reach an equilibrium state by minimizing the total energy. The complete hysteresis loop was reconstructed by assuming symmetric behavior when sweeping the field from +800 to -800 mT.
Finally, the magnetoresistance was calculated using a normalized version of Eq. (\ref{DeGennes}).

\medskip
\textbf{Acknowledgements} \par 

The authors would like to thank Abhishek Naik and Nicolas Lejeune for their valuable input in the MFM and FMR measurements, and Koen Schouteden for his useful advise regarding the AFM topography scans. The authors acknowledge financial support from Fonds de la Recherche Scientifique - FNRS and Flemish Research Foundation - FWO, under the grant Weave PDR T.0208.23-FNRS and G0D7723N-FWO, and by COST (European Cooperation in Science and Technology) [www.cost.eu] through COST Action SUPERQUMAP (CA 21144). S.R. acknowledges funding by Fonds Wetenschappelijk Onderzoek (grant nr. 11A3V25N). The work of E.F. has been partially supported by the FWO and F.R.S.-FNRS under the Excellence of Science (EOS) project O.0028.22. J.V.d.V  acknowledges support from the KU Leuven C14/21/083 C1-program. J.Y.G. acknowledges the funding by National Key Research and the Science and Technology Commission of Shanghai Municipality (Grant No. 24CL2910702). B.D. acknowledges the Fonds de la Recherche Scientifique (FRS-FNRS), B-1000 Bruxelles, Belgium. Simulation time was provided by the Consortium d'Equipements de Calcul Intensif (FRS-FNRS Belgium Grant No. 2.5020.11).

\medskip
\newpage
\bibliographystyle{MSP}
\bibliography{Ref.bib}

\pagebreak
\medskip

\appendix
\newpage

\title{Cryogenic Voltage Control of Magnetism in Silicon-Integrated \newline SrTiO$_3$/Fe Heterostructures : Supplementary Information}

\maketitle

\author{Stijn Reniers$^{ 1}$*}
\author{Emile Fourneau$^{ 2}$}
\author{Andries Boelen$^{ 3,4}$}
\author{Xing-Jian Liu$^{ 5}$}
\author{Ekaterina Gorokh$^{ 1}$}
\author{Lukas Nulens$^{ 1}$}
\author{Vivek Kumar$^{ 1}$}
\author{Luca Ceccon$^{ 3,4}$}
\author{Christian Haffner$^{ 4}$}
\author{Clement Merckling$^{ 3,4}$}
\author{Jun-Yi Ge$^{ 5}$*}
\author{Bertrand Dup\'e$^{ 6}$}
\author{Alejandro V. Silhanek$^{ 2}$}
\author{Kristiaan Temst$^{ 1,4}$}
\author{Joris Van de Vondel$^{ 1}$*}

\begin{affiliations}
	
$^{1}$ Quantum Solid-State Physics, Department of Physics and Astronomy, KU Leuven, Celestijnenlaan 200D, Leuven B-3001, Belgium
	
$^{2}$Experimental Physics of Nanostructured Materials, Q-MAT research unit, Department of Physics, Université de Liège, Liège B-4000, Belgium
	
$^{3}$Department of Materials Engineering (MTM), KU Leuven, B-3001 Leuven, Belgium
	
$^{4}$Imec, Kapeldreef 75, Leuven, Belgium
	
$^{5}$Materials Genome Institute, Shanghai University, Shanghai 200444, China
	
$^{6}$TOM research group, Q-MAT research unit, Université de Liège, Liège B-4000, Belgium
	
*stijn.reniers@kuleuven.be
*joris.vandevondel@kuleuven.be
*junyi\_ge@t.shu.edu.cn
\end{affiliations}

\keywords{Voltage Control of Magnetism, Cryogenic Nanomagnetism, Spin Reorientation Transition}

\subsection{Growth and structural characterization}

Detailed information on the epitaxial growth of the Strontium Titanate film can be found in the Experimental Methods section. In this section, we provide additional specific information on the epitaxial overgrowth of the ultrathin Fe layer. 

\begin{figure*} [h]
	\centering
	\includegraphics[width=7in]{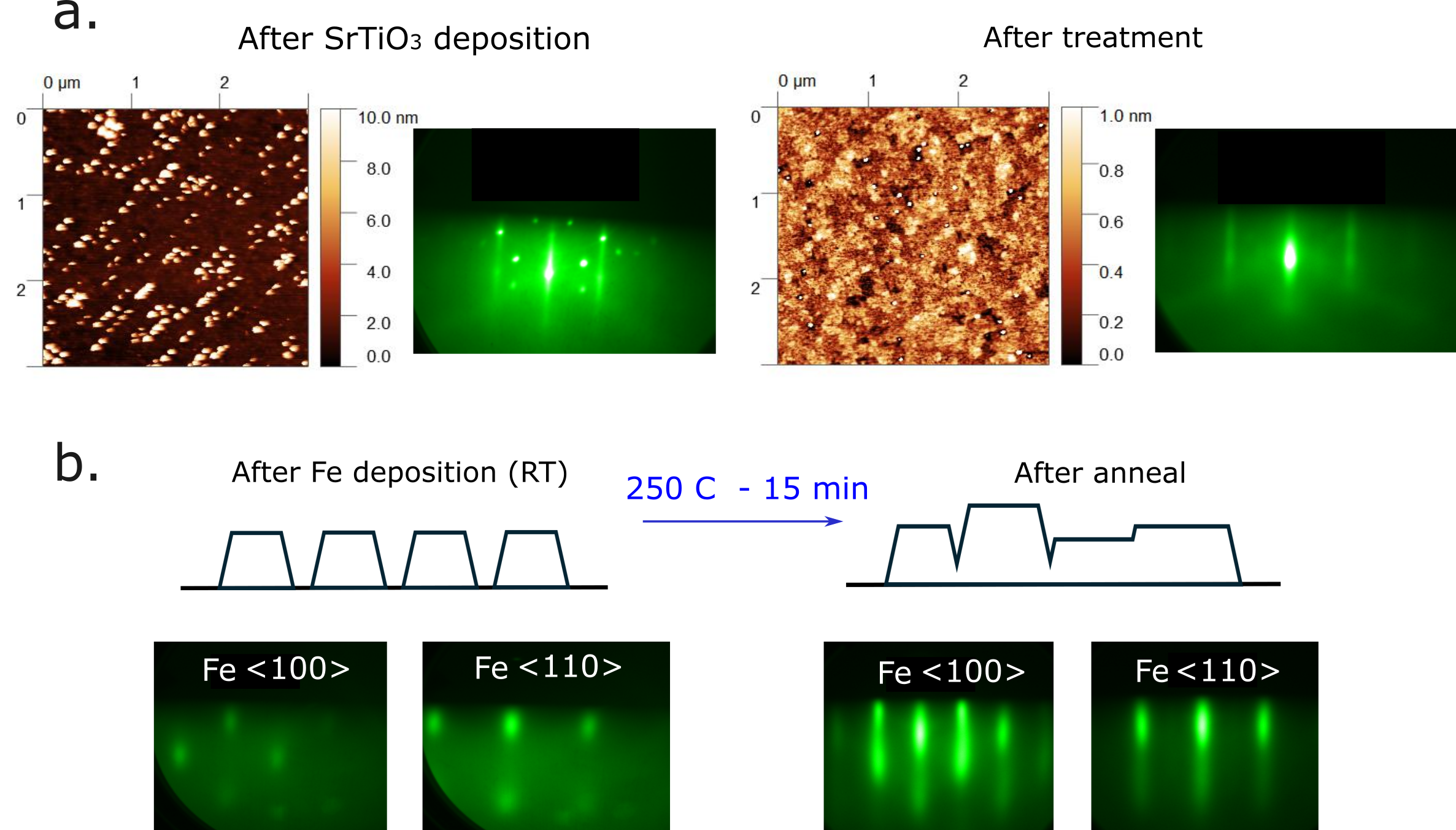}
	\caption{(a) Effect of BHF etch + annealing treatment on SrTiO$_3$ thin film surface. (b) Deposition protocol of the Fe ultrathin layer. Evolution of the RHEED-pattern from separated spots to streaks indicates coalescence of the Fe islands after in-situ annealing. }
	\label{fig:DepoFe}
\end{figure*}

Prior to loading the samples in the second UHV chamber after SrTiO$_3$ film deposition, they were subjected to a chemical treatment, consisting of a 20 minute sonication in deionized water, followed by a 30 second immersion in buffered HF solution (Experimental methods). Once in the UHV, the samples were outgassed briefly (15 minutes) at 300 degrees Celsius for removal of residual water and hydrocarbon contaminants, and then left to cool back to room temperature. It is well-known that below a few nm thickness, Fe tends to grow in Volmer-Weber growth \cite{Fenanocrystals} on SrTiO$_3$. To maintain a continuous Fe layer, we first deposit at room temperature, to limit the mobility of diffusing Fe atoms on the SrTiO$_3$ surface. This creates a network of small and densely spaced islands. Following this first step, the samples are annealed at 250 Celsius for 15 minutes, which causes coalescence of neighboring islands and enhances the crystallinity of the Fe film. The resulting film is not perfectly flat, but structurally continuous and epitaxial, which is sufficient to support robust ferromagnetism and a clean Fe/SrTiO$_3$ interface. As such, it fulfills the requirements for VCM studies on this system.

\vspace{10 pt}

\begin{figure*} [h!]
	\centering
	\includegraphics[width=6.5in]{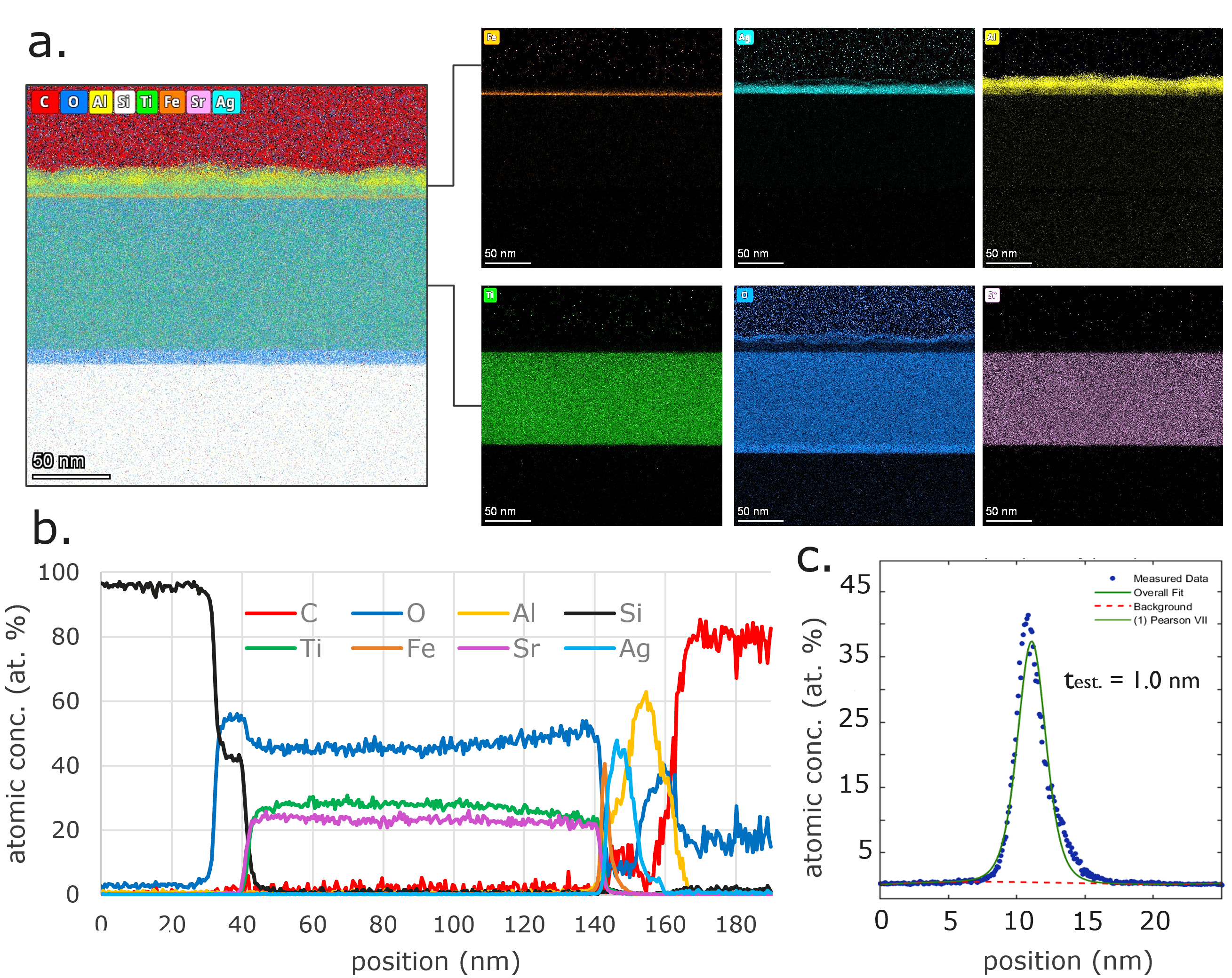}
	\caption{ (a) EDX characterization of the full thin film heterostructure, showing the defined elemental profiles of each layer and continuity of the Fe film. (b) EDX line profile acquired along the heterostructure, showing the average elemental profile. (c) Pearson fit of the Fe EDX peak, used for a rough nominal thickness estimate.}
	\label{fig:TEMEDX}
\end{figure*}

In addition to the region near the Fe/SrTiO$_3$ interface, EDX measurements were performed for the full thin film stack. Figure \ref{fig:TEMEDX}a shows the EDX compositional map, with the individual layer contributions in separate insets. The compositional profile is given in Figure \ref{fig:TEMEDX}b. It shows an interfacial SiO$_2$ layer (thickness 8.5 $\pm$ 1 nm) and the slightly Ti-rich composition of the SrTiO$_3$ film. To verify the Fe nominal thickness $t_{Fe}$, the Fe concentration profile was fitted using a Pearson distribution curve in LIPRAS peak fitting software, integrated and divided by 100\%. This gives an effective thickness of 1.0 nm, which is above the nominal thickness $t_{Fe} = 0.8$ nm, obtained using quartz crystal calibration during deposition. The observed difference could attributed to finite probe size or interface broadening effects, which are significant at sub-nanometer lengthscales for EDX.

\pagebreak 

\subsection{Volumetric magnetometry measurements}

\begin{figure*}[h!]
	\centering
	\includegraphics[width=7 in]{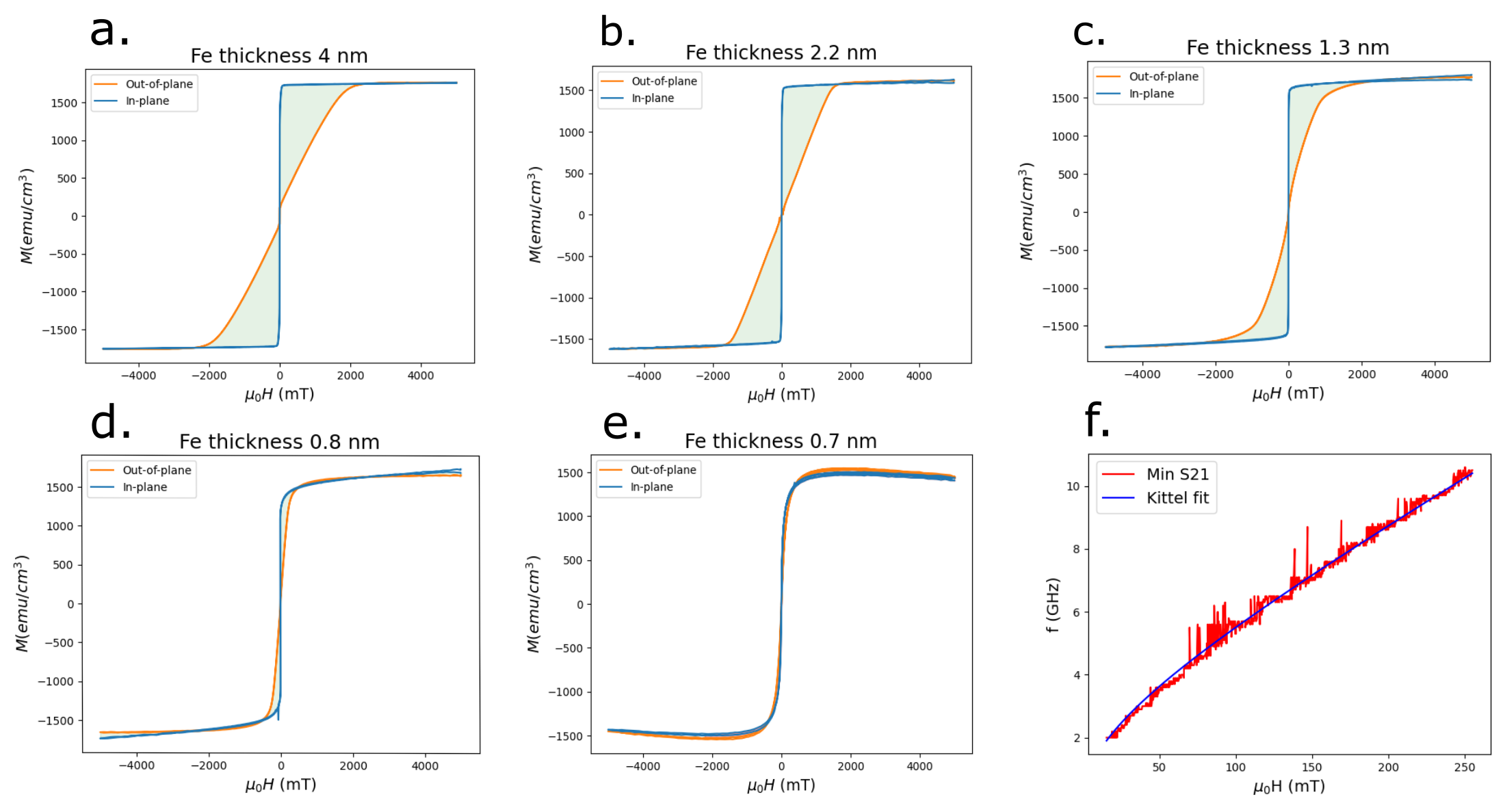}
	\caption{(a-e) SQUID-VSM magnetometry measurements for Fe thicknesses $t_{Fe}$ = 4 nm, 2.2 nm, 1.3 nm, 0.8 nm and 0.7 nm. The effective anisotropy can be derived from the magnetizing work difference as $K_{eff} = W_{OOP}-W_{IP}$, which corresponds to half of the shaded area. (f) Kittel fit to the extracted $S_{21}$ minimum from the FMR measurement at $T$ = 300 K.}
	\label{fig:supp_squid}
\end{figure*}

The thickness dependence of the effective areal magnetic anisotropy $K_{eff} \cdot t$ was found by analysis of SQUID-VSM measurements, performed in a Quantum Design MPMS3 system. After hand-cleaving the samples by a diamond tip, the Fe volume was estimated using the sample area, multiplied by the nominal thickness. The absolute magnetization $M$ was obtained by division of the measured magnetic moment by the estimated Fe volume, after linear subtraction of the diamagnetic background, originating from the Si substrate. Figure \ref{fig:supp_squid}a-e shows the obtained $M(H)$-loops for the magnetic field applied in- and perpendicular to the sample plane. The work density $W$ required to magnetize the Fe sample along the field axis, is given by :

\begin{equation}
	W = \mu_0 \int_{0}^{Ms}{HdM} 
\end{equation}

The magnetizing work was computed for both in-plane and out-of-plane magnetization curves using trapezoidal integration, for different Fe film thicknesses, yielding $K_{eff} = W_{OOP}-W_{IP}$. This was done for the ascending and descending branches, and the average value was used. The error bars on the volume, magnetization and effective anisotropy energy density were computed as :

\begin{equation}
	\delta V =  \sqrt{((\frac{\delta A}{A})^2 + (\frac{\delta t}{t})^2)V^2} \space \rightarrow \space \delta M =  \frac{\delta V}{V} M \space\rightarrow \space \delta K_{eff} \approx  \frac{\delta V}{V}K_{eff} \space
\end{equation}

The in-plane FMR resonance line at 300 K was fitted using the modified Kittel equation (Figure \ref{fig:supp_squid}f):
\begin{equation}
	f_{res}(H_0) = \frac{\gamma}{2\pi}\sqrt{H_0(H_0+(M_s-H_k))} = \frac{\gamma}{2\pi}\sqrt{H_0(H_0+(M_s-\frac{2K_{u}}{M_s}))}
\end{equation}

Where $\gamma$ is the gyromagnetic ratio (28 GHz/T) and $H_K = \frac{2K_{u}}{\mu_0M_s}$ is the uniaxial interface anisotropy field, originating from the room temperature uniaxial interface anisotropy $K_u = \frac{K_{int}}{t_{Fe}}$. We fit the parameter $M_s - H_k$ = 0.3 T. Using a value of $M_s$ = 1770 $\pm$ 243 kA/m, obtained from SQUID-VSM measurements, we find $H_K = 1.9 \pm 0.3$ T, equivalent to a value of $K_u$ = 1.7 $\pm$ 0.2 MJ/m$^3$ or $K_{int}$ = 1.3 $\pm$ 0.2 mJ/m$^2$  .

\subsection{Density Functional Theory}

\begin{figure*}[h!]
	\centering
	\includegraphics[width=7 in]{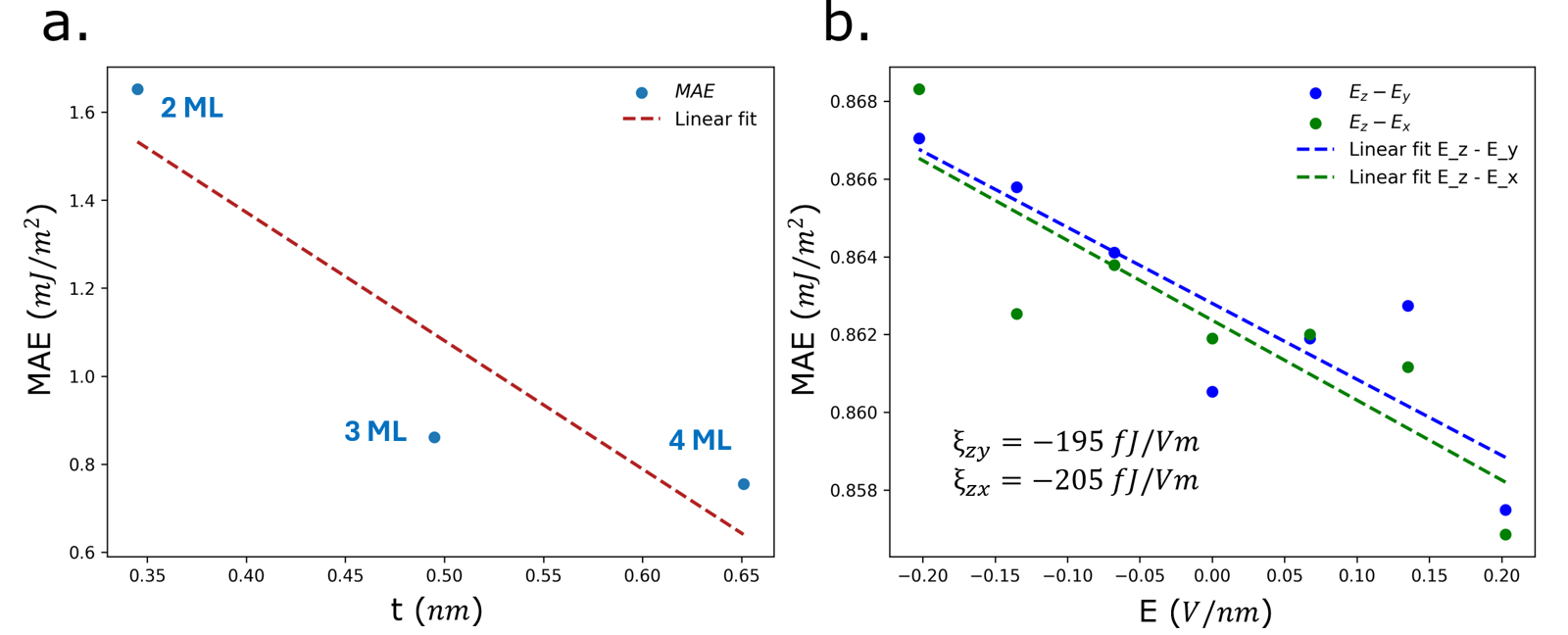}
	\caption{(a) Thickness dependence of the MAE-value for different numbers of Fe monolayers in the DFT heterostructure. Linear fit is used to extrapolate the interfacial contribution. (b) First-principles calculations of the VCMA coefficient $\xi$ as the difference of the total SOC contributions to the total energy $E_z-E_y$ and $E_z-E_x$, respectively. Linear fitting provides an first-order estimate of $\xi$.}
	\label{fig:supp_DFT}
\end{figure*}

Figure \ref{fig:supp_DFT}a shows the Magnetic Anisotropy Energy (MAE), as the first-principles equivalent to the experimental areal magnetic anisotropy energy density $K_{eff} \cdot t$, calculated for SrTiO$_3$/Fe/Ag heterostructures with different number of Fe monolayers (ML = 2,3,4). Spin-orbit coupling (SOC) provides the main source of MAE, so the MAE is computed as the difference of total SOC contribution to the total energy between the out-of-plane z-(001) and in-plane x-direction (100). While the results for 3 and 4 ML are relatively close to the experimental trend, it can be seen that the 2 ML MAE value is substantially enhanced. This potentially causes a strong distortion to the linear extrapolation of $K_{int}$ towards higher values, overestimating the $K_{int}$-value. 
The electric field-dependence of MAE was calculated for the 3 monolayer Fe slab, shown in Figure \ref{fig:supp_DFT}b. The change of interfacial SOC with electric field is the main mechanism of electronic VCM or VCMA. We computed the MAE relative to both in-plane orientations x and y, as shown in Figure \ref{fig:supp_DFT}b. The obtained, first order VCMA coefficient $\xi$ is negative, which is consistent with our experimental observations. Furthermore, the obtained values are in the same order of magnitude as our experimentally estimated value from magnetotransport measurements. It should be noted that, since we do not take into account low-temperature structural and dielectric properties of SrTiO$_3$, these results are not fully representative of the VCM in our stack. Theoretical work focusing specifically on the low-temperature regime would be a valuable continuation of this work. 

\subsection{Magnetotransport measurements} 

\subsubsection{Transverse Magnetoresistance and exchange scattering model}

\begin{figure*}[h!]
	\centering
	\includegraphics[width=7 in]{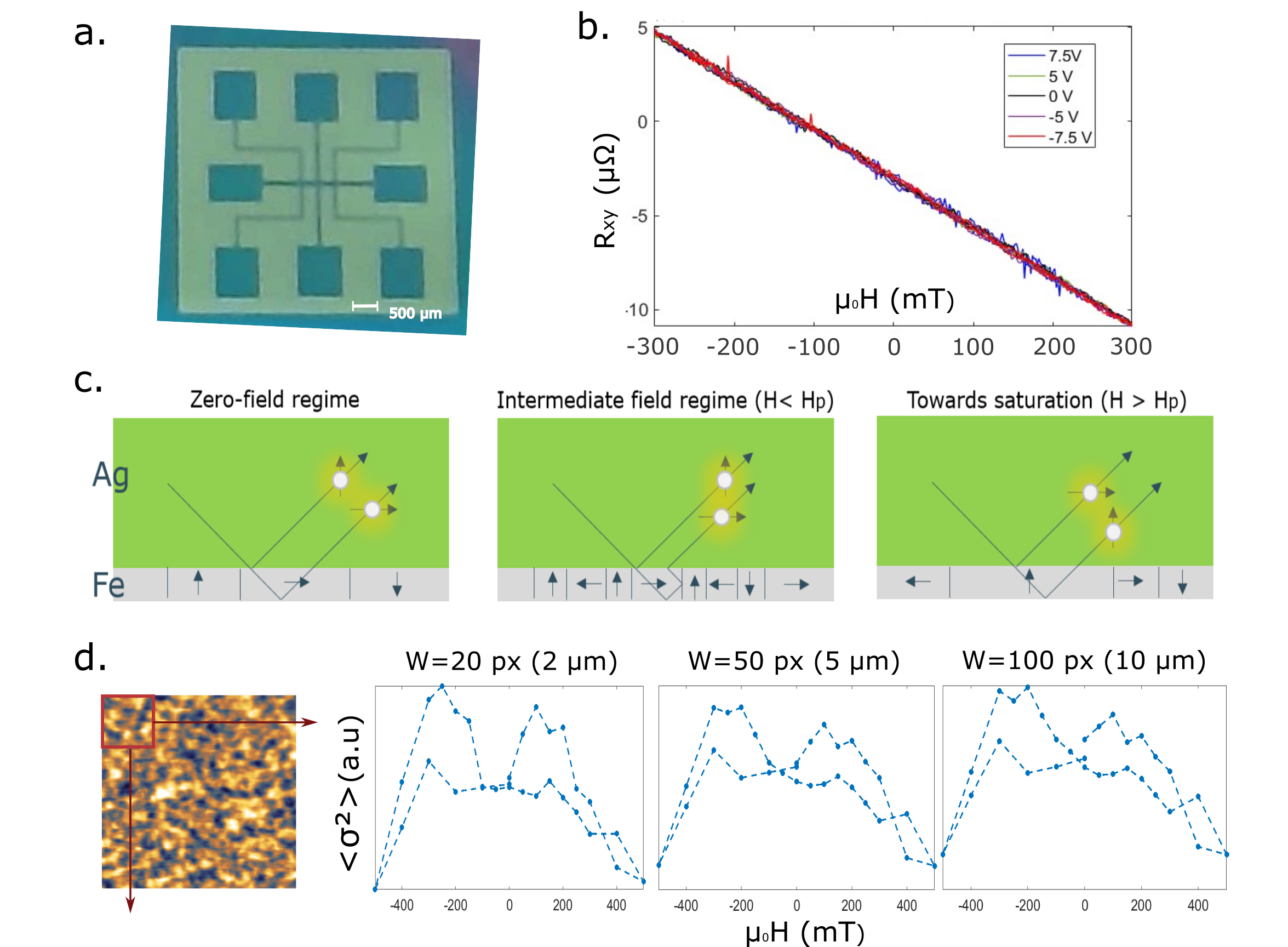}
	\caption{(a) Optical image of the patterned Hall bar sample, used in the magnetotransport measurements. (b) Transverse magnetotransport measurements at different gate voltage. The linear dependence indicates a regular Hall effect, with a small offset voltage, potentially originating from slight contact misalignment. (c) Schematic illustration of the proposed scattering mechanism underlying the magnetotransport observations. (d) Extraction procedure for $<\sigma^2>$ out of the MFM images of 15 x 15 $\mu$m$^2$ size. A sliding window is used to compute the variance in phase value over the window area. The average is then computed for all window positions.}.
	\label{fig:supp_tran}
\end{figure*}

Figure \ref{fig:supp_tran}a shows an optical image of the Hall bar transport structure sample used for magnetotransport measurements. In Figure \ref{fig:supp_tran}b,  the transverse resistance $R_{xy}$ as function of applied magnetic field shows a strictly linear dependence, originating from the regular Hall effect. The absence of any anomalous contribution indicates that transport is dominated by spin-dependent scattering at the Fe/Ag interface rather than coherent orbital scattering in the Fe film.
The physical picture relating the scattering mechanism to the longitudinal magnetoresistance $R_{xx}(H)$ is schematically illustrated in Figure \ref{fig:supp_tran}c. In the field regime below domain nucleation, the mean free path of conduction electrons in the Ag layer is shortened as they interact with Fe domains of different spin orientation, with a scattering probability directly related to the spin-disorder. As the magnetic field increases, nucleated domains grow by pinning-limited domain wall motion. While structural defects act as effective pinning sites, we are mainly interested in the pinning caused by local magnetic anisotropy variations. This pinning enhances spin disorder and hence the overall scattering probability of the conduction electrons, increasing the longitudinal resistance. At high fields, beyond the peak field $H_p$ in magnetoresistance, the effective pinning barrier is overcome and domain walls can propagate more smoothly and domains grow larger. This reduces again the overall scattering probability and related resistance.

The analysis of MFM images, which we use to correlate our magnetotransport observations to the magnetic microstructure, was performed by image analysis in Matlab, using the \textit{imread}() functionality. The procedure is shown in Figure \ref{fig:supp_tran}d.  A sliding window was used to compute the local variance in color intensity along all 3 RGB-channels, then computes the moving average of that variance. The sliding window size for Figure 3 in the main text was 20 x 20 pixels (2 x 2 $\mu$m). While a smaller sliding window gives a more locally-resolved disorder, larger window sizes yield similar qualitative behavior, which reproduces the general trend of the $R_{xx}(H)$ curve. 

\subsubsection{VCMA-coefficient : derivation from general and microscopic arguments} \label{VCMA derivation}

Based on insights from earlier works \cite{KIRILYUK199745, DW_pinning_PRB, DW_pinning_PRL} and our micromagnetic simulations, we argue that local regions with stronger uniaxial interface anisotropy $K_u(x,y) = K_{int}/t(x,y)$ play an dominant role in the domain wall pinning process. Due to their limited thickness, these regions will experience a larger voltage-induced shift in $K_u(x,y)$ compared to thicker regions with low or negative (in-plane favored) anisotropy. To reason how this is reflected in the observed transport behavior, we assume that the field $H_p$ relates to the average effective pinning barrier energy $\Delta_{DW}$ as:

\begin{equation}
	\mu_0H_pM_s = \Delta_{DW} + \epsilon
\end{equation}

\begin{figure*}[h!]
	\centering
	\includegraphics[width=7 in]{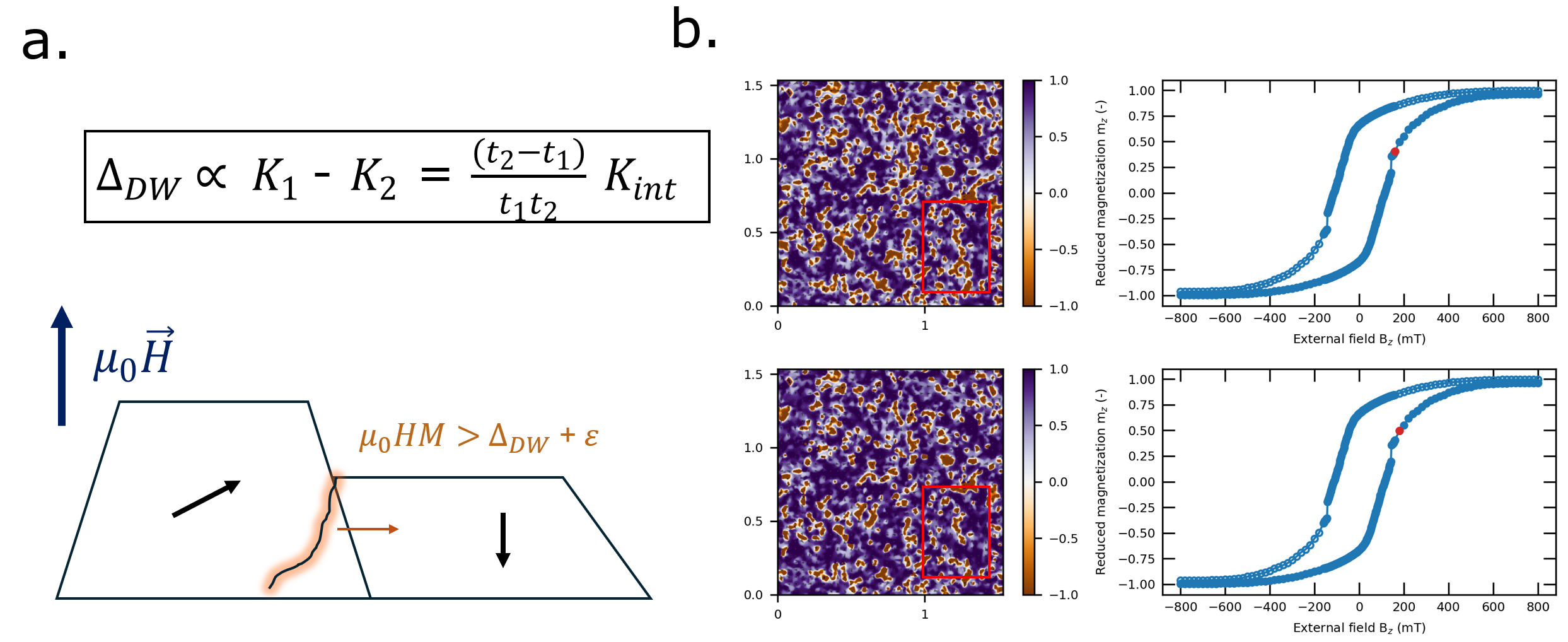}
	\caption{(a) Schematic illustration of the domain wall pinning process at the boundaries of different anisotropy regions. The component of the pinning barrier related to anisotropy variation scales with the variation amplitude $\Delta K$. (b) Micromagnetic illustration of 'depinning' at a field beyond the coercive field. Red boxes indicate a region where strong perpendicular anisotropy regions switch orientation.}.
	\label{fig:supp_VCM}
\end{figure*}

This relation represents the point where the Zeeman energy overcomes the average pinning potential. The main physical question to produce an effective VCM-coefficient estimate $\xi$, lies in the link between the interface magnetic anisotropy and the value of $\Delta_{DW}$. Figure \ref{fig:supp_VCM}a gives a schematic representation of the anisotropy contribution to the energy barrier. Since the domain wall needs to move to a region of higher magnetic anisotropy, it will cause an energy cost proportional to the differences in local anisotropies $\Delta_{DW} \propto K_1-K_2$. A micromagnetic illustration of the pinning effect is shown in Figure \ref{fig:supp_VCM}b.

The proportionality factor is key for the estimate of the VCM-coefficient, and is strongly related to the anisotropy contrast and \textit{pinning correlation lengthscale parameter $\zeta$}, reflecting how strongly the pinning landscape fluctuates compared to the domain-wall width.

\begin{equation}
	\mu_0M_s\frac{\delta H_p}{\delta E} =  \frac{\delta \Delta_{DW}}{\delta E} = \zeta \cdot\frac{\delta K_{Fe/SrTiO_3}\Delta t_{Fe}}{t_{Fe}^2\delta E} = \eta \frac{\delta K_{Fe/SrTiO_3}}{\delta E}
\end{equation}

Where $\eta = \zeta \cdot\frac{\Delta t_{Fe}}{t_{Fe}^2}$. Since we have strong pinning, and thickness variations occur on a scale of $\approx$ 10 nm, we are in the regime where $\zeta$ is in the order of 1. For thickness fluctuations in the order of 25 to 50 \% around the mean, the $\frac{\Delta t_{Fe}}{t_{Fe}^2}$ is smaller than 1. We can then produce an estimate of the VCM coefficient, estimating $\eta \approx \mathcal{O}$(1) :
\begin{equation}
	\xi = \frac{\delta K_{Fe/SrTiO_3}}{\delta E} \approx \frac{\mu_0\delta H_pM_st_{Fe}}{\delta E} 
\end{equation} 

Filling in the values $\delta H_p$ = 12 mT, $\delta E$ = 0.15 V/nm, $M_s \approx 1700$ kA/m and $t_{Fe}$ = 0.8 nm, we obtain $\xi$ = - $\mathcal{O}$(100) $\frac{fJ}{Vm}$. It should be stressed that this value is a linear approximation, based on overcoming of the pinning energy barrier caused by local regions of high magnetic anisotropy.  

\subsection{Magneto-Optic Kerr Effect Microscopy measurements} 

\begin{figure*}[h!]
	\centering
	\includegraphics[width=7 in]{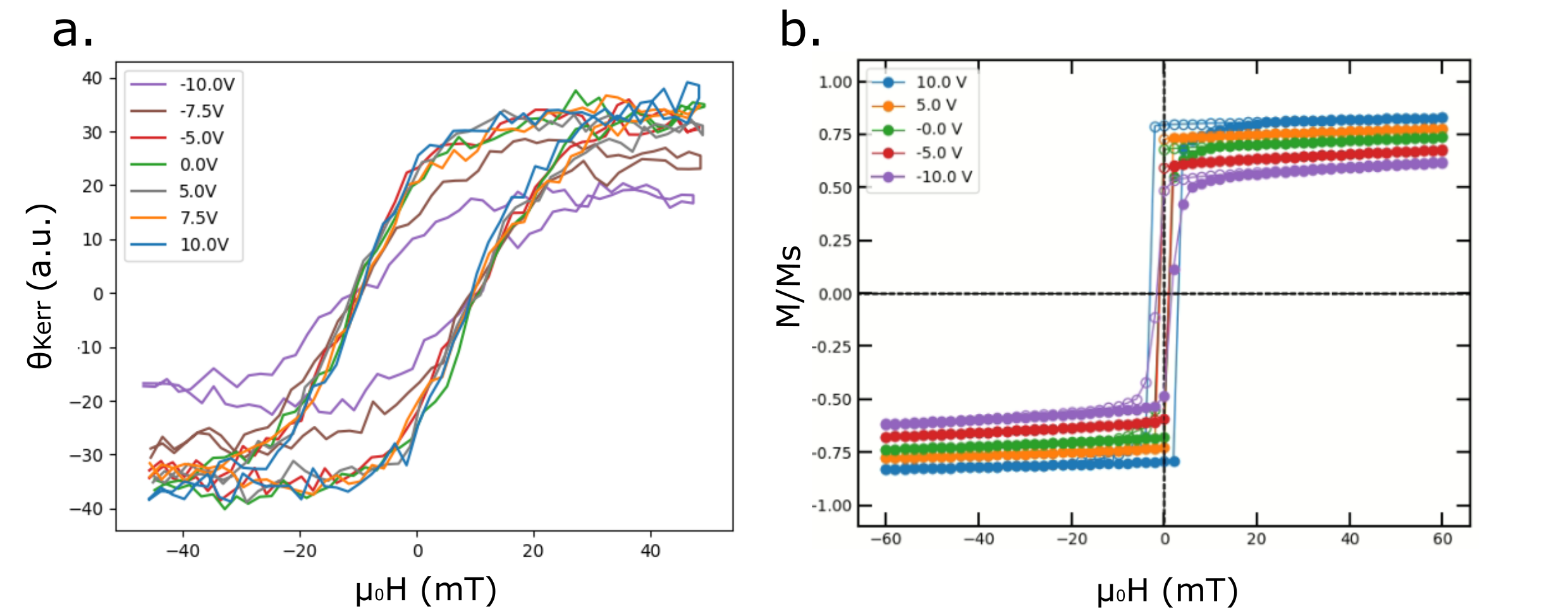}
	\caption{(a) Experimentally measured LMOKE hysteresis loops at $T$ = 5 K, for different back-gate voltages. (b) Micromagnetic simulations for reduced magnetization hysteresis loops for different back gate voltages.}.
	\label{supp:MOKE}
\end{figure*}

Figure \ref{supp:MOKE}a shows the measured LMOKE hysteresis loops at $T$ = 5 K for all applied back-gate voltages. It can be seen that the evolution of loop area and magnetic contrast is non-linear with gate voltage. The non-linearity is much more limited in the micromagnetic simulations of reduced magnetization as a function of field (Figure \ref{supp:MOKE}b). However, the overall trend and polarity dependence of gate voltage is the same. The much sharper switching at reversal and lower coercive field suggest that the simulation slab does not fully capture the domain pinning for in-plane magnetic fields.

\subsection{Magnetic Force Microscopy measurements} 

\begin{figure*}[h!]
	\centering
	\includegraphics[width=7 in]{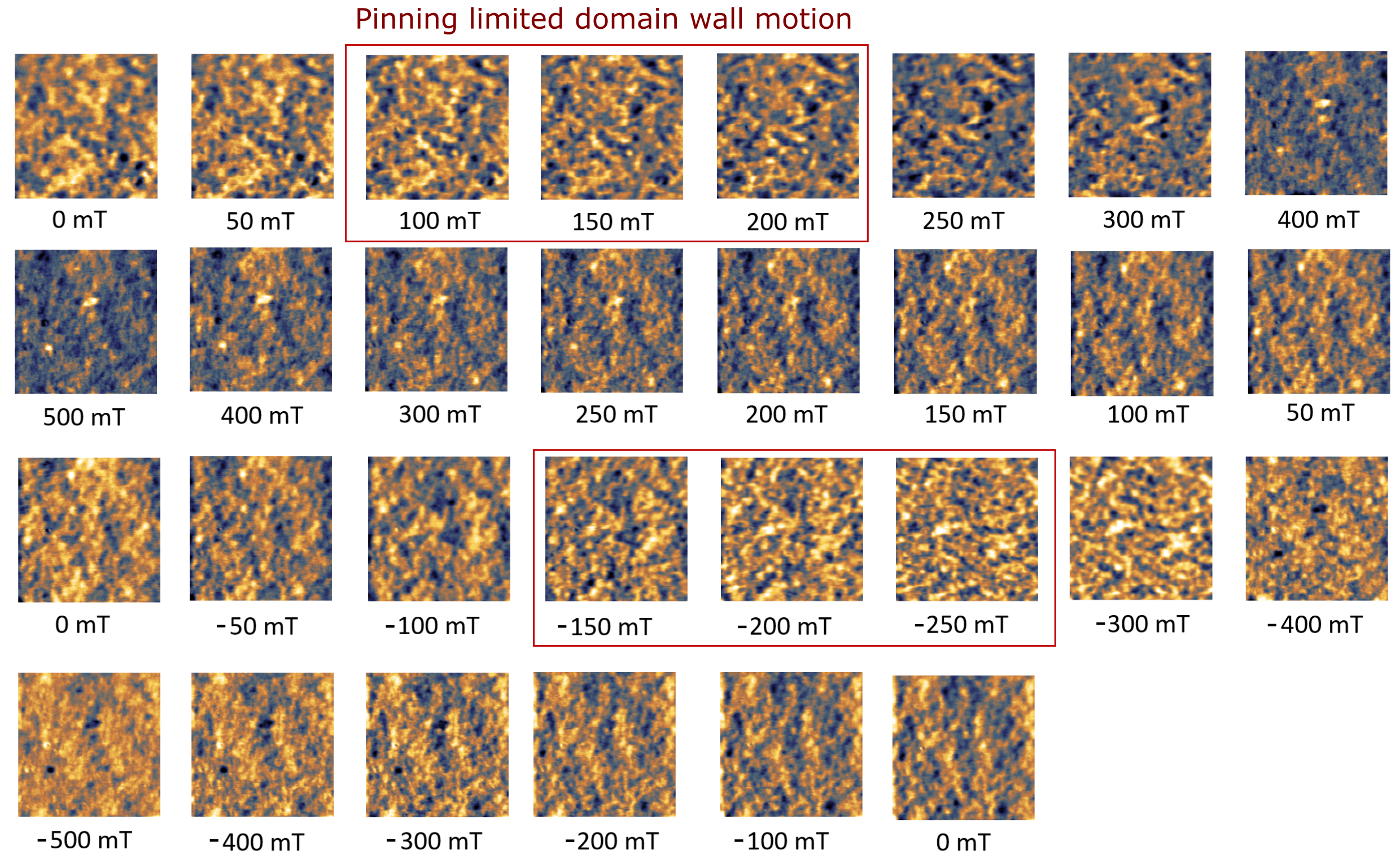}
	\caption{Magnetic Force Microscopy images (15 x 15 $\mu$m), recorded during the same magnetic field sequence as the transport measurement at 0V back-gate voltage. Red boxes indicate the pinning-limited domain wall motion regime, resulting in the highest values of magnetoresistance.}
	\label{fig:supp_MFM1}
\end{figure*}

MFM-images shown in Figure \ref{fig:supp_MFM1} were measured in the pristine sample state at scan range 15 x 15 $\mu$m$^2$ under application of varying magnetic field during a M(H)-loop within 500 mT range. A fine domain structure can be identified for all images, with maximal reduction in the coherent domain size for the intermediate field ranges, indicated in the red boxes. The regions correspond roughly to the peak region in magnetotransport measurements, and represent the pinning limited domain wall propagation regime. Domains deform as their boundaries try to overcome local pinning potentials, enhancing their local irregularity and global spin disorder.

\begin{figure*}[h!]
	\centering
	\includegraphics[width=6 in]{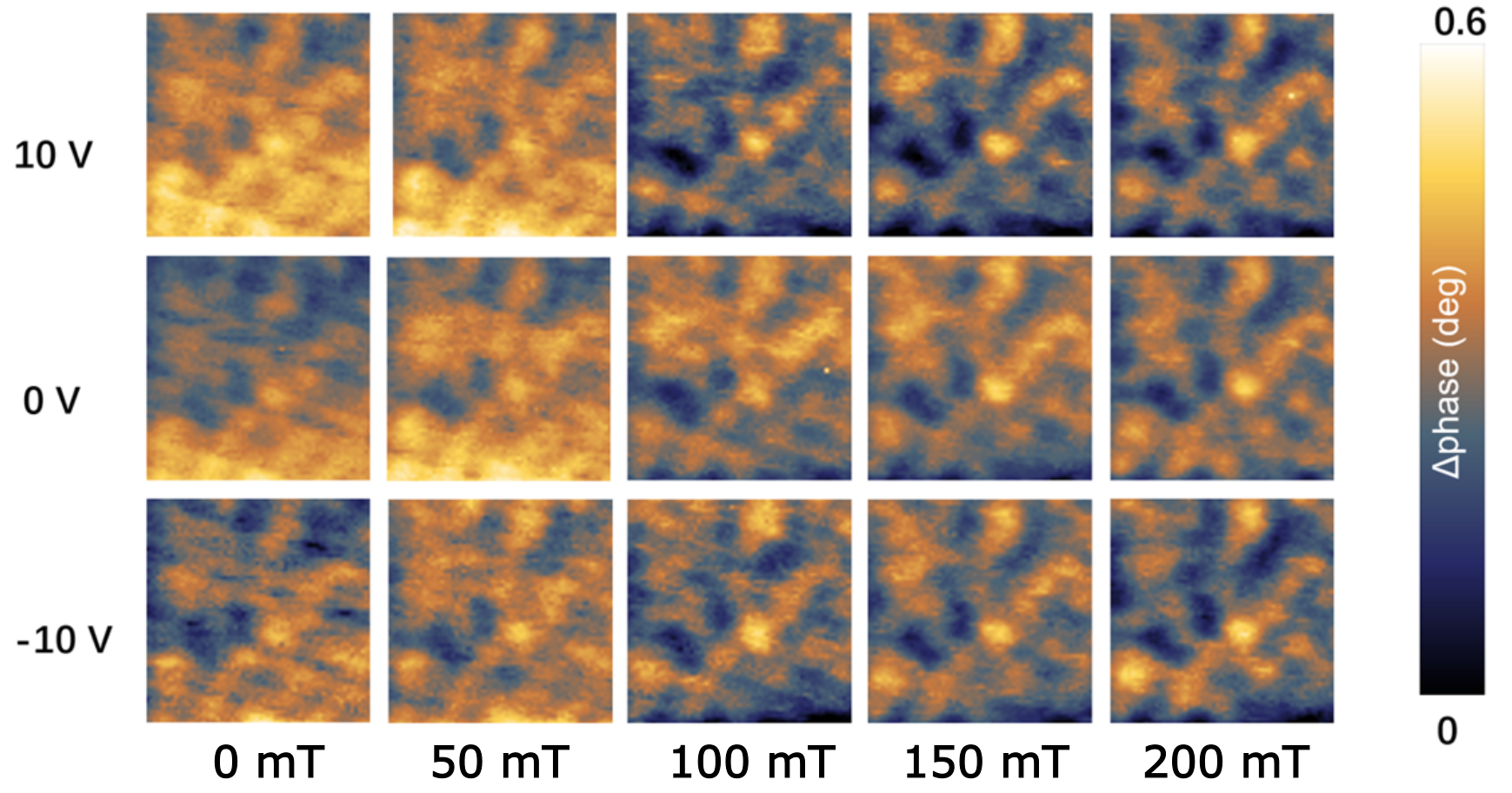}
	\caption{Magnetic Force Microscopy images (4 x 4 $\mu$m), recorded at back gate voltages of -10 V, 0 V and 10 V.}
	\label{fig:supp_MFM2}
\end{figure*}

\vspace{10 pt}
As an attempt to capture the VCM effect on the smallest microscopic scale, MFM images were recorded in a reduced scan range of 4 x 4 $\mu$m$^2$, under the application of back gate voltages  -10 V, 0 V and 10 V (Figure \ref{fig:supp_MFM2}). While the gate effect is hard to distinguish clearly, some useful general observations can be made. The larger phase contrast for -10 V at 0 mT indicates a stronger out-of-plane component, as expected for negative gate voltage. Furthermore, the regime starting from 100 mT shows a more dispersed domain structure, which does not change drastically upon increase of the magnetic field. This is caused by the domain wall pinning. Based on our physical arguments, the domain structure should become more disordered as the field approaches $H_p$, but it is hard to distinguish this within the scale of the measurement. This disorder enhancement is shown more clearly for the larger scale images of Figure \ref{fig:supp_MFM1}. Within the pinning-limited regime, the gate effect, correlated with shifts of the magnetotransport peaks, is not clearly distinguishable.  We conclude that, although an impact is visible in the remanent state, the MFM images recorded for out-of-plane magnetic field are not the ideal method for clear visualization of the back-gate effect. We therefore choose to focus on the MOKE-microscopy results of the main text, performed with in-plane magnetic field or in magnetic remanence.

\end{document}